\documentclass[twocolumn,preprintnumbers,pra,superscriptaddress,showkeys]{revtex4}%smaller,showpacs,

\usepackage{amsmath,amssymb,amsfonts,graphicx,float,times,psfrag}
\usepackage[pdftex]{color}
\usepackage{amsmath,bm}
\usepackage[colorlinks, linkcolor=blue, citecolor=blue,  breaklinks=true]{hyperref}
\usepackage{mathtools,amsfonts,mathptmx}
\usepackage[utf8]{inputenc}
\usepackage[T1]{fontenc}
\usepackage{xcolor}
\usepackage{braket}
\usepackage[titletoc,title]{appendix}
\usepackage[nameinlink,capitalize]{cleveref}

\usepackage[shortlabels]{enumitem}
\begin{document}

\title{Multi-path vector entanglement engineering via dark mode control in optomechanics} 
\author{P. Djorwé}
\email{djorwepp@gmail.com}
\affiliation{Department of Physics, Faculty of Science, 
University of Ngaoundere, P.O. Box 454, Ngaoundere, Cameroon}
\affiliation{Stellenbosch Institute for Advanced Study (STIAS), Wallenberg Research Centre at Stellenbosch University, Stellenbosch 7600, South Africa}

\author{R. Altuijri}
\email{raaltuwagry@pnu.edu.sa}
\affiliation{Department of Physics, College of Science, Princess Nourah bint Abdulrahman University, P.O. Box 84428, Riyadh 11671, Saudi Arabia}

\author{A. J. Almalki}
\email{alialmalki@ub.edu.sa}
\affiliation{Department of Computer Science and Artificial Intelligence, College of Computing and Information Technology, University of Bisha, Bisha 61922, Saudi Arabia}

\author{S. Abdel-Khalek}
\email{asbotalb@tu.edu.sa}
\affiliation{Department of Mathematics and Statistics, College of Science, Taif University, P.O. Box 11099, 21944, Saudi Arabia}

\author{A.-H. Abdel-Aty}
\email{amabdelaty@ub.edu.sa}
\affiliation{Department of Physics, College of Sciences, University of Bisha, Bisha 61922, Saudi Arabia}

% \author{F. Kunst}
% \email{flore.kunst@mpl.mpg.de}
% \affiliation{Max Planck Institute for the Science of Light, 91058 Erlangen, Germany}
% \affiliation{Department of Physics, Friedrich-Alexander-Universität Erlangen-Nürnberg, 91058 Erlangen, Germany}

\begin{abstract}
We propose a scheme to generate multi-paths entanglement in an optomechanical system by exploiting polarized electromagnetic fields and dark mode control. Our system consists of two mechanically coupled mechanical resonators, which are driven by a common electromagnetic field. An inclusion of a polarizer induces linear polarizations of the electromgnetic field corresponding to the vertical (transverse electric ($\rm{TE}$) and horizontal (transverse magnetic [($\rm{TM}$]) modes, which drive the mechanical resonators. Without the mechanical coupling $J_m=0$, the polarization angle ($\phi$) controls dark mode in the system. The breaking of this dark mode leads to multi-paths engineering of bipartite optomechanical entanglements.  By switching on the phonon hopping rate ($J_m\neq0$), both the polarization angle and the modulation phase of the mechanical coupling allow a further control of the dark mode. The simultaneous Dark Mode Breaking (\rm{DMB}) conditions under these two parameters leads to multi-paths bipartite and tripartite entanglements. For a fine tuning of the polarization angle ($\phi=\pi/4$) this scheme enables a generation of twin entangled states, where the bipartite/tripartite generated entangled states are degenerated and might be of great interest for quantum information processing, quantum communication and diverse quantum computational tasks. The generated entanglements are more resilient against thermal fluctuations in the \rm{DMB} regime, i.e., up to two order of magnitude robust than in the Unbreaking regime.  Our work sheets light on new  possibilities to generate noise-tolerant quantum resources that are useful for plethora of modern quantum technologies. 
\end{abstract}

\maketitle
\date{\today}

\section{Introduction}\label{intro}
Quantum entanglement is an interesting and  vital resource for modern quantum technologies including quantum information processing \cite{Slussarenko2019,Wendin2017}, quantum metrology \cite{Pezz2018,Polino2020} and sensing \cite{Degen2017,Djorwe2019,Djorw2024,Tchounda2023}, quantum communication protocols \cite{Stannigel2010,Knaut2024,Kucera2024}, and quantum computational tasks \cite{McArdle2020,Bharti2022,Bartolucci2023}, just to name few. However, generating such non-classical states is not an easy task, since they require specific engineering schemes for their stabilization \cite{Shah2024,Chen.2025} against thermal fluctuations and other external environmental perturbations during their processing. Several way of generating stable an robust entangled states have been proposed in different fields, for instance in  plasmonic systems \cite{Tame2017}, electron spin systems \cite{Hensen2015}, superconducting circuits \cite{Steffen2006,DiCarlo2010,Palomaki2013}, and optomechanics \cite{Chen2025}. 

Different optomechanical structures have  been recently used as benchmark systems to generate quantum entanglements.  Optomechanical entanglement, i.e., entanglement between electromagnetic field and center of mass of a mechanical object has been firstly investigated in \cite{Vitali2007}. Later on, other interesting works on entanglement involving optomechanical systems and hybrid related systems have been realized both theoretically and experimentally. Owing to their vulnerability against environmental disturbances,  entangled states are fragile for certain applications. To handle such an issue, number of studies on macroscopic entanglement have been realized. A stabilization of macroscopic entanglement of two massive micromechanical oscillators have been achieved in \cite{Korppi2018},  where the mechanical oscillators were coupled to a microwave-frequency electromagnetic cavity. A pulsed electromechanical system has led to a direct entanglement observation of two mechanical drumheads through quantum state tomography in \cite{Kotler2021}. Similarly, a Duan quantity of $1.4$ decibels below the separability bound has been measured \cite{Mercier2021}, revealing quantum entanglement of two micromechanical oscillators. Beside of these interesting achievements, there are recent theoretical proposals aiming to enhance quantum entanglement. For instance, different  nonlinearities  have been used to improve entanglement such as Duffing nonlinearity \cite{Massembele2024}, Kerr nonlinearity \cite{Yang2020}, Brillouin scattering \cite{Djo2024,Rostand2025},  and quadratic coupling \cite{Ghorbani2025}. Moreover, sophisticated technics based on exceptional points \cite{Li2023} and dark mode control have been proposed as well \cite{Lai2022,Lai.2022}. Despite these aforementioned entanglement investigations in optomechanical structures, the urge to propose schemes that generate abundant entanglement resources is still crucial for applications involving large network of entities. Such schemes can be useful to overcome the entanglement sharing among multiple users, which remains a fundamental issue in modern quantum computational tasks. For this purposes, multipath quantum entanglement sources have been proposed \cite{Rossi2009,Zhang2023}, enabling an efficient distribution of the quantum resources among multiuser of a multiplexed architecture \cite{Man2023,Sutcliffe2023}. While some of these protocols are based on a generation of two photons by spontaneous parametric down-conversion \cite{Rossi2009,Hu_2020}, and other based  on segment routing/quantum repeaters \cite{Zhang2023}, recent developments in optomechanics make use of polarization of electromagnetic field.

Polarization of electromagnetic field has been used to foster interesting phenomena in physics owing to the fact that the polarized fields provide extra free control parameter to the system. The induced phenomena include Faraday effect \cite{Duggan2019}, chiral exceptional point polarization conversion \cite{Oh2024}, and entanglement \cite{Bowen2002}, among others. In optomechanical systems, polarizations of electromagnetic field have been used to perform optomechanically induced transparency \cite{Xiong2016}, phonon laser \cite{Wang2018}, and quantum entanglement \cite{Buters2016,Li2021}. The entanglement polarization have been investigated so far in single  optomechanical cavity (where the  polarized fields interact with a mechanical resonator), and it only focuses on bipartite entanglement. In our proposal, we extend this concept to a large structure  where the polarized fields interact with two mechanical resonators, and we investigated both bipartite and tripartite quantum entanglements. Our proposal can be seen as an optical cavity driven by linear polarizations of the electromgnetic field, i.e., the vertical (transverse electric ($\rm{TE}$) and horizontal (transverse magnetic [($\rm{TM}$]) modes, which interact with two mechanical resonators placed within the cavity. The involved mechanical resonators are mechanically coupled through a phonon hopping rate $J_m$, which is $\theta-$modulated. When the phonon hopping rate is off ($J_m=0$), we have used the polarization angle to control dark mode in the system, and this has led to a generation of multi-path bipartite entanglements. In this case, no tripartite entanglement was engineered. As the mechanical coupling is switched on  ($J_m\neq0$),  both the polarization angle ($\phi$) and the modulation phase ($\theta$) are used to control dark mode, leading to interesting outcomes. Under the simultaneous Dark Mode Breaking (\rm{DMB}) conditions, i.e., realized by tuning both $\phi$ and $\theta$, we have found that i) multi-paths bipartite and tripartite entanglements are generated, providing abundant quantum resources, ii) twin bipartite/tripartite entangled states are engineered for $\phi=\pi/4$, and iii) the generated entanglements are up to two order of magnitude robust against thermal noises compared to those engineered in the Unbreaking regime. Our proposal constitutes a rich platform to generate noise-tolerant quantum resources which are of great interest for quantum information processing, quantum communication network and modern quantum computational tasks. 

The rest of the work is organized as follow. The model and the involved equations are described in \autoref{sec:model}, together with the theoretical quantification of the entanglements. Bipartite and tripartite entanglements are investigated in \autoref{sec:bipart} and \autoref{sec:tripart}, respectively. Our conclusion is presented in \autoref{sec:Concl}.

\section{Model and dynamical equations} \label{sec:model}
Our benchmark system consists of a linearly polarized electromagnetic field whose polarized components simultaneously drive two mechanical resonators. Both mechanical resonators are mechanically coupled through a phonon hopping rate $J_m$. This mechanical coupling is modulated through a phase $\theta$, which can be used to induces a synthetic magnetism is the system \cite{Mathew2020,Slim2025}. In a frame rotating with respect to the driving frequency $\omega_L$, the Hamiltonian of our system (with $\hbar=1$) is given by, 
\begin{equation}
 H=H_O+H_{OM}+H_{int}+H_{drive},
\end{equation}
with
\begin{equation}
\begin{cases}
 H_{O}&=-\sum_{p=\updownarrow,\leftrightarrow} \Delta_p a_p^\dagger a_p + \sum_{j=1,2} \omega_j b_j^\dagger b_j, \\
 H_{OM}&= \sum_{j=1,2} \left[ \sum_{p=\updownarrow,\leftrightarrow} g_p a_p^\dagger a_p (b_j^\dagger + b_j) \right], \\
H_{int}&=J_m \left(e^{i\theta}b_1^\dagger b_2 + e^{-i\theta} b_1 b_2^\dagger \right),\\
H_{drive}&=i\sqrt{\kappa} \sum_{p=\updownarrow,\leftrightarrow} \alpha_p^{in} (a_p^\dagger + a_p),
\end{cases}
\end{equation}
where $H_O$, $H_{OM}$, $H_{int}$, and $H_{drive}$ are the Hamiltonians for the free optomechanical system, the optomechanical coupling, the mechanical interaction, and the driving fields, respectively. The double arrows $\updownarrow,\leftrightarrow$ denote the linear polarizations of the electromagnetic field driving the mechanical resonators.  We have defined the detuning $\Delta_p=\omega_L-\omega_c^p$, with $\omega_c^p$ being the  cavity frequency related to the $p^{th}$ polarized field. The electromgnetic (mechanical) modes are captured by the annihilation $a_p$ ($b_j$) and creation $a_p^\dagger$ ($b_j^\dagger$) operators. The mechanical frequencies are denoted by $\omega_j$, while $g_j$ stands for the single-photon optomechanical coupling rate. 

\begin{figure}[tbh]\label{fig:fig1}
\begin{center}
  \resizebox{0.45\textwidth}{!}{
  \includegraphics{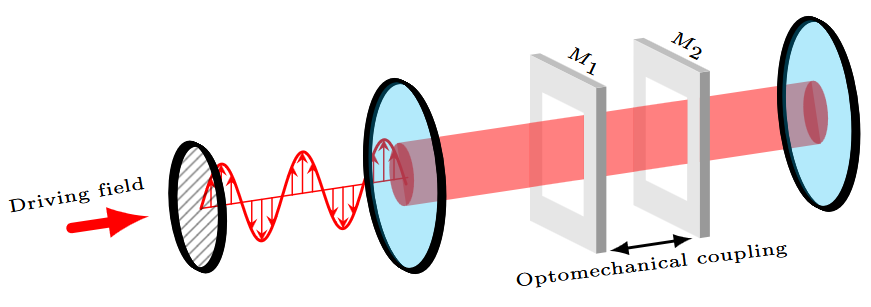}}
  \end{center}
\caption{Sketch of our proposal. Two mechanically coupled mechanical resonators are placed inside an optical resonator, which is driven by polarized electromagnetic fields.}
\label{fig:Fig1}
\end{figure}

\begin{figure}[tbh]
\begin{center}
  \resizebox{0.45\textwidth}{!}{
  \includegraphics{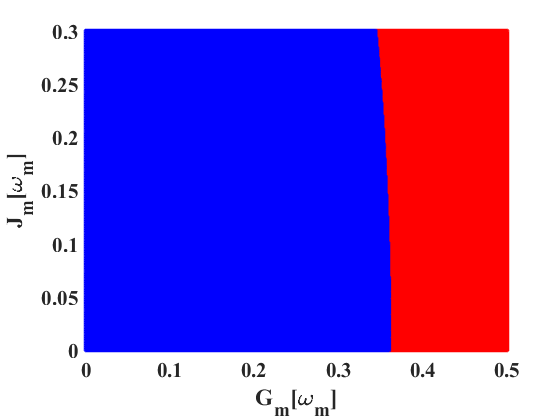}}
  \end{center}
\caption{Stability diagram of the system depending on the effective coupling $G_m$ and the mechanical coupling $J_m$. The blue/dark region is stable, while the red/light area represents the unstable zone. The used parameters are $\tilde{\Delta}=-\omega_m$, $\kappa=0.2\omega_m$, $\omega_j=\omega_m$, $\gamma_j=10^{-5}\omega_m$, $\phi=\frac{\pi}{4}$, and $\theta=\frac{\pi}{2}$.}
\label{fig:fig2}
\end{figure}

The amplitude of the linearly polarized driving field is $\alpha^{in}=|\alpha_\updownarrow^{in}|^2+|\alpha_\leftrightarrow^{in}|^2$, where $\alpha_\updownarrow^{in}=\alpha^{in}\cos{\phi}$ and $\alpha_\leftrightarrow^{in}=\alpha^{in}\sin{\phi}$ being the projections of $\alpha^{in}$ onto the vertical and horizontal modes, respectively. The driving field is related to the input power $P^{in}$ through $\alpha^{in}=\sqrt{P^{in}/\hbar \omega_L}$. The aforementionned linear polarizations of the electromgnetic field correspond to the vertical (transverse electric [TE]) and horizontal (transverse magnetic [TM]) modes of the Fabry–Pérot cavity. These linearly polarized light can be thought as a superposition of orthogonal patterns encoded into the unit vector $|\textbf{e}\rangle= \cos{\phi} |\textbf{e}_\updownarrow\rangle + \sin{\phi}|\textbf{e}_\updownarrow\rangle$, where $\phi$ is the angle between the linearly direction of the involved polarization. In such a configuration, one can coherently manipulate the spatial distribution of optical field by tunning  its polarization structure through the angle $\phi$. By invoking the Heisenberg equation, one can derive the following Quantum Langevin Equations (QLEs),
\begin{equation}\label{QLE}
\begin{cases}
\dot{a}_p&=\left\{ i\left[\Delta_p - g_p\sum_{j=1}^2 (b_j^\dagger + b_j)\right] - \frac{\kappa}{2} \right\}a_p +\sqrt{\kappa}\alpha_p^{in}  ,\\
\dot{b}_j&=-\left(i\omega_j + \frac{\gamma_m}{2} \right)b_j -iJ_m b_{3-j}e^{(-1)^{3-j}i\theta} - i \sum_{p=\updownarrow,\leftrightarrow} g_pa_p^\dagger a_p.
\end{cases}
\end{equation}
where we have assumed the same dissipations for the polarized modes ($\kappa_{\updownarrow}\equiv\kappa_{\leftrightarrow}\equiv \kappa$), and for the mechanical modes ($\gamma_1\equiv\gamma_2\equiv \gamma$). From now on, we will also assume that the two polarized modes couple to the mechanical resonators with the same rate ($g_{\updownarrow}\equiv g_{\leftrightarrow}\equiv g$). By using the standard linearization procedure, where the operators are splitted into their mean value and some amount of fluctuations, i.e., $a_p=\alpha_p + \delta \alpha_p$ and $b_j=\beta_j + \delta \beta_j$, one gets the mean dynamical equations,
\begin{equation}
\begin{cases}
\dot{\alpha}_p&=\left( i\tilde{\Delta}_p - \frac{\kappa}{2} \right)\alpha_p +\sqrt{\kappa}\alpha_p^{in} ,\\
\dot{\beta}_j&=-\left(i\omega_j + \frac{\gamma_m}{2} \right)\beta_j -iJ_m \beta_{3-j}e^{(-1)^{3-j}i\theta} - ig \sum_{p=\updownarrow,\leftrightarrow} |\alpha_p|^2,
\end{cases}
\end{equation}
and the fluctuations dynamics,
\begin{equation}\label{eq:fluc}
\begin{cases}
\delta\dot{\alpha}_p&=\left(i\tilde{\Delta}_p  - \frac{\kappa}{2} \right)\delta \alpha_p - ig \alpha_p \sum_{j=1} (\delta \beta_j^\dagger + \delta\beta_j) + \sqrt{\kappa}\delta\alpha_p^{in} ,\\
\delta\dot{\beta}_j&=-\left(i\omega_j + \frac{\gamma_m}{2} \right)\delta\beta_j -iJ_m \delta\beta_{3-j}e^{(-1)^{3-j}i\theta} \\&- ig \sum_{p=\updownarrow,\leftrightarrow} \alpha_p(\delta\alpha_p^\dagger+ \delta\alpha_p) + \sqrt{\gamma_m}\delta\beta_j^{in},
\end{cases}
\end{equation}
with the effective detuning $\tilde{\Delta}_p=\Delta_p-2g\sum_{j=1} \rm Re(\beta_{j})$. We have also defined the effective couplings $G_p=g\alpha_p$, i.e., $G_{\updownarrow}=G_m\cos{\phi}$ and $G_{\leftrightarrow}=G_m\sin{\phi}$, with $G_m=g\alpha_0$ corresponding to $\phi=0$. The quantum fluctuations are captured by  $\delta\alpha_p^{in}$, and $\delta\beta_j^{in}$ which are zero-mean noise operators characterized by the following auto-correlation functions,
\begin{eqnarray}\label{eq:noise}
\langle \delta\alpha_p^{in}(t)\delta\alpha_p^{in\dagger}(t') \rangle =&\delta(t-t'), \nonumber \\ 
\langle \delta\alpha_p^{in\dagger}(t)\delta\alpha_p^{in}(t') \rangle =& 0, \nonumber \\
\langle \delta\beta_j^{in}(t)\delta\beta_j^{in\dagger}(t') \rangle  =& (n_{th}^j+1)\delta(t-t'), \nonumber \\
\langle \delta\beta_j^{in\dagger}(t)\delta\beta_j^{in}(t') \rangle =& n_{th}^j\delta(t-t'),\nonumber 
\end{eqnarray}
where $n_{th}$ is the thermal phonon occupation of the mechanical resonator defined as $n_{th}=[\rm exp(\frac{\hbar \omega_m}{k_bT})-1]^{-1}$, where $\rm k_b$ is the Boltzmann constant.\\
To quantify the bipartite entanglement within our system, we need to compute the covariance matrix through the quantum quadratures. For this purpose, we introduce the following quadratures, $\delta I_p=\frac{1}{\sqrt{2}}(\delta \alpha_p + \delta \alpha_p^\dagger)$, $\delta Y_p=\frac{i}{\sqrt{2}}(\delta \alpha_p^\dagger - \delta \alpha_p)$, $\delta X_j=\frac{1}{\sqrt{2}}(\delta \beta_j + \delta \beta_j^\dagger)$ and $\delta P_j=\frac{i}{\sqrt{2}}(\delta \beta_j^\dagger - \delta \beta_j)$, together with their corresponding noise quadratures. The substitution of these quadratures into the fluctuation equations leads to the quadrature equations, which can be cast into their following compact form,

\begin{equation}\label{eq:qua}
\delta\dot{X}=\rm{M}\delta X+ \rm{N} \delta X^{in}
\end{equation}
with the column vector $\delta X=(\delta{I_{\leftrightarrow}},\delta{Y_{\leftrightarrow}},\delta{I_{\updownarrow}},\delta{Y_{\updownarrow}}, \delta{X}_1,\delta{P}_1, \delta{X}_2,\delta{P}_2)^T$, and the corresponding noise column vector $\delta X^{in}=(\delta{I}_{\leftrightarrow}^{in},\delta{Y}_{\leftrightarrow}^{in},\delta{I}_{\updownarrow}^{in},\delta{Y}_{\updownarrow}^{in}, \delta{X}_1^{in},\delta{P}_1^{in}, \delta{X}_2^{in},\delta{P}_2^{in})^T$. The drift matrix $N$ is given by ${\rm N}=Diag(\sqrt{\kappa},\sqrt{\kappa},\sqrt{\kappa},\sqrt{\kappa},\sqrt{\gamma_m} ,\sqrt{\gamma_m},\sqrt{\gamma_m},\sqrt{\gamma_m})$. In the above quadrature equations, the matrix $M$ reads,  

\begin{widetext}
\begin{equation}
\rm{M}=\begin{pmatrix}
-\frac{\kappa}{2}&-\tilde{\Delta}_{\leftrightarrow}&0&0&2\rm Im(G_{\leftrightarrow})&0&2\rm Im(G_{\leftrightarrow})&0 \\ 

\tilde{\Delta}_{\leftrightarrow}&-\frac{\kappa}{2}&0&0&-2\rm Re(G_{\leftrightarrow})&0&-2\rm Re(G_{\leftrightarrow})&0\\ 

0&0&-\frac{\kappa}{2}&-\tilde{\Delta}_{\updownarrow}&2\rm Im(G_{\updownarrow})&0&2\rm Im(G_{\updownarrow})&0 \\ 

0&0&\tilde{\Delta}_{\updownarrow}&-\frac{\kappa}{2}&-2\rm Re(G_{\updownarrow})&0&-2\rm Re(G_{\updownarrow})&0\\

2\rm Im(G_{\leftrightarrow})&0&2\rm Im(G_{\updownarrow})&0&-\frac{\gamma_m}{2}&\omega_1 & J_m\sin\theta & J_m \cos\theta\\ 

-2\rm Re(G_{\leftrightarrow})&0&-2\rm Re(G_{\updownarrow})&0&-\omega_1 &-\frac{\gamma_m}{2}& -J_m \cos\theta & J_m \sin\theta \\ 

2\rm Im(G_{\leftrightarrow})&0&2\rm Im(G_{\updownarrow})&0&-J_m \sin\theta &J_m \cos\theta &-\frac{\gamma_m}{2}&\omega_2 \\

-2\rm Re(G_{\leftrightarrow})&0&-2\rm Re(G_{\updownarrow})&0&-J_m \cos\theta & -J_m\sin\theta &-\omega_2 &-\frac{\gamma_m}{2}
\end{pmatrix}.
\end{equation}
\end{widetext}
From the above quadrature equations, one can now define the covariance's matrix elements as  $V_{ij}=\frac{\langle \delta X_i \delta X_j  + \delta X_j \delta X_i \rangle}{2}$, which also satisfy the motional equation,
\begin{equation}\label{eq:lyap1}
\dot{V}={\rm M}V+V{\rm M^T}+D,
\end{equation}
where the diagonal diffusion matrix is expressed as $D=Diag[\frac{\kappa}{2},  \frac{\kappa}{2},\frac{\kappa}{2},  \frac{\kappa}{2}, \frac{\gamma_m}{2}(2n_{th}^1 + 1), \frac{\gamma_m}{2}(2n_{th}^1 + 1), \frac{\gamma_m}{2}(2n_{th}^2 + 1), \frac{\gamma_m}{2}(2n_{th}^2 + 1)]$.  To be meaningful, the matrix $\rm M$ must satisfy the Routh-Huritz stability criterion, i.e., all its eigenvalues should have negative real parts \cite{DeJesus1987}. Our used parameters have been chosen wihin this stability range as depicted in \autoref{fig:fig2}. This figure displays the stability of our system versus the phonon hopping rate $J_m$ and the effective coupling $G_m$. The blue/dark region captures the system parameters enabling stability, while the red/light area represents the unstable zone.

To quantify the steady-state behavior of the entanglement, we consider the long time limit of the dynamics in \autoref{eq:lyap1}, where the variable are no longer time dependent. Therefore, the motional equation in \autoref{eq:lyap1} reduces to the Lyaponuv equation,

\begin{equation}\label{eq:lyap}
{\rm M}V+V{\rm M^T}=-D,
\end{equation}
and the covariance matrix takes the general form, 
\begin{equation}\label{cov}
\rm{V}=
\begin{pmatrix} 
V_{\alpha_\leftrightarrow}&V_{\alpha_{\leftrightarrow},\alpha_{\updownarrow}}&V_{\alpha_{\leftrightarrow},\beta_1}&V_{\alpha_{\leftrightarrow},\beta_2}  \\
V_{\alpha_{\leftrightarrow},\alpha_{\updownarrow}}^{\intercal}&V_{\alpha_{\updownarrow}}&V_{\alpha_{\updownarrow},\beta_1} &V_{\alpha_{\updownarrow},\beta_2}\\ 
V_{\alpha_{\leftrightarrow},\beta_1}^{\intercal}&V_{\alpha_{\updownarrow},\beta_1}^{\intercal}&V_{\beta_1}&V_{\beta_1,\beta_2}\\
V_{\alpha_{\leftrightarrow},\beta_2}^{\intercal}&V_{\alpha_{\updownarrow},\beta_2}^{\intercal}&V_{\beta_1,\beta_2}&V_{\beta_2}
\end{pmatrix},
\end{equation}
where $V_{\mu}$ and  $V_{\mu\nu}$ are blocks of $2\times2$ matrices (with $\mu,\nu \equiv \alpha_p, \beta_j$). The covariance matrix elements can be analytically evaluated, but this leads to a tedious task, and to cumbersome expressions. Therefore, these expressions will be computed numerically instead. For simplicity and without loss of the generality, we assume that our mechanical resonators are degenerated, i.e., $\omega_j\equiv\omega_m$ and $\gamma_j\equiv\gamma_m$. We have used the following art-of-the-state optomechanical parameters \cite{Fang2017,Brendel2017,Mathew2020,Slim2025}, i.e., $ \omega_m/2\pi=10 \rm{MHz}$; $\omega_1=\omega_m$,  $\omega_2=\omega_m$, $\kappa=0.2\omega_m$, $\Delta_p=-\omega_m$, $\gamma_m=10^{-5}\omega_m$, $G_m=0.2\omega_m$, and $J_m=0.2\omega_m$. The phase $\theta$ and the angle  $\phi$ are adjusted latter on.

\begin{figure}[tbh]
\begin{center}
  \resizebox{0.45\textwidth}{!}{
  \includegraphics{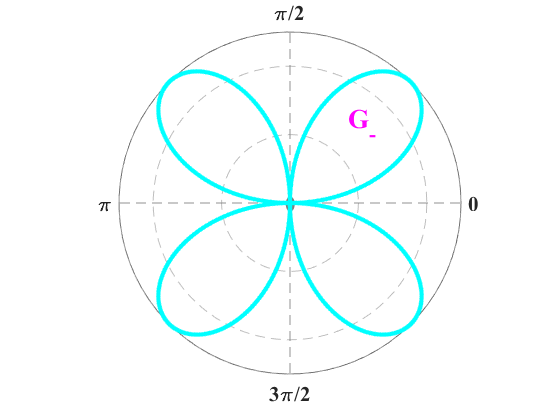}}
  \end{center}
\caption{(a)  Coupling strength $G_-$ in polar-coordinate versus the polarized angle $\phi$ for $G_m=0.2\omega_m$, $\tilde{\Delta}_{\updownarrow}=-\omega_m$, and $\tilde{\Delta}_{\leftrightarrow}=-(1+10^{-3})\omega_m$. The other parameters are $J_m=0$, $\kappa=0.2\omega_m$, $\omega_j=\omega_m$, and $\gamma_j=10^{-5}\omega_m$.}
\label{fig:fig3}
\end{figure}

The diagonal blocks denoted as $V_{\mu}$ correspond to the optical modes ($\mu = \alpha_p$), and the mechanical mode ($\mu= \beta_j$), respectively. The off-diagonal blocks capture the correlations between different subsystems, i.e., $V_{\alpha_p,\beta_j}$ describes the correlations between the $p^{th}$ intracavity field and the $j^{th}$ mechanical mode. A bipartite entanglement within two subsystems can be quantified using the logarithmic negativity ($E_N$), which can be evaluated by tracing out the non-necessary modes. The logarithmic negativity $E_N$ is defined as \cite{Vidal2002,Plenio2005},
\begin{equation}\label{eq:en}
E_N=\rm max[0,-\ln(2\nu^-)], 
\end{equation}
where $\nu^- =\rm{min}$ $\rm{eig} |i\Omega_2 \tilde{\chi}_4|$, with the symplectic matrix $\Omega_2=\otimes^2_{j=1}i\sigma_y$, and $\sigma_y$ being the $y-$Pauli matrix. In this definition,  $\nu^-$ refers to the minimum symplectic eigenvalue of the covariant matrix $\tilde{\chi}_4=P_{1|2}\chi_4 P_{1|2}$, where $\chi_4$ is a $4\times4$ matrix of the targeted subsystems that is defined as,
\begin{equation}
\rm{\chi_4}=\begin{pmatrix} 
V_{\mu}&V_{\mu\nu} \\
V_{\mu\nu}^{\intercal}&V_{\nu}
\end{pmatrix},
\end{equation}
and $P_{1|2}=\rm{diag}(1,-1,1,1)$ is a matrix that realizes partial transposition at the level of $\chi_4$. From \autoref{eq:en}, there is an entanglement emerging from the system if and only if the condition $\nu^-<1/2$ is fulfilled, which is equivalent to Simon's necessary and sufficient entanglement criterion for Gaussian states.  Throughout the work, we consider the red-sideband detuning driving for the mechanical resonator ($\tilde{\Delta}_p=-\omega_m$), where sufficient heating processes are suppressed in the system.

In order to quantify tripartite entanglement,  we compute the minimum residual contangle between each polarized mode and the two mechanical resonators. Therefore, we define $R^{\alpha_h|\beta_1|\beta_2}$ ($R^{\alpha_v|\beta_1|\beta_2}$) as the minimum residual contangle between $\rm{TM}$ ($\rm{TE}$) mode with the two mechanical resonators $m_{j=1,2}$. For a given triplet  of mode $u,v,w$, the  minimum residual contangle is defined as \cite{Adesso2006,Adesso2007}, $R^{u|v|w} =  R^{u|(vw)}-R^{u|v}-R^{u|w}$, ($u,v,w \equiv \alpha_{h,v}, \beta_1, \beta_2$). In this formulation, $R^{u|v}$ stands for the contangle of subsystems labeled by $u$ and $v$ (where $v$ can contain one or two modes), which is the squared logarithmic negativity. To evaluate the one-mode-vs-two-modes logarithmic negativity $R^{u|(vw)}$, one just needs to consider the definition of \autoref{eq:en} where we substitute the quantities  $\Omega_2=\otimes^2_{j=1}i\sigma_y$ and $\tilde{\chi}_4=P_{1|2}\chi_4 P_{1|2}$  with $\Omega_3=\otimes^3_{j=1}i\sigma_y$ and $\tilde{\chi}_6=P_{u|vw}\chi_6 P_{u|vw}$, respectively. In these new definitions, $\chi_6$ is $6\times6$ covariance matrix for the subsystem of interest, $P_{1|23}=\rm{diag}(1,-1,1,1,1,1)$, $P_{2|13}=\rm{diag}(1,1,1,-1,1,1)$ , and $P_{3|12}=\rm{diag}(1,1,1,1,1,-1)$ are partial transposition matrices. To be a proper entanglement measure, the residual contangle must fulfill the monogamy of quantum entanglement, i.e., $R^{u|v|w}>0$ that is similar to the Coffman-Kundu-Wootters monogamy inequality \cite{Coffman2000}. Therefore, the quantification of CV tripartite entanglement is, $R^{min} \equiv \rm{min} \left[ R^{\alpha_{h,v}|\beta_1|\beta_2},R^{\beta_1|\alpha_{h,v}|\beta_2},R^{\beta_2|\alpha_{h,v}|\beta_1}\right]$, which ensures that the residual contangle  $R^{min}$ is invariant under all permutations of the modes resulting in a genuine three-way property of any three-mode Gaussian state. 

\begin{figure}[tbh]
\begin{center}
  \resizebox{0.45\textwidth}{!}{
  \includegraphics{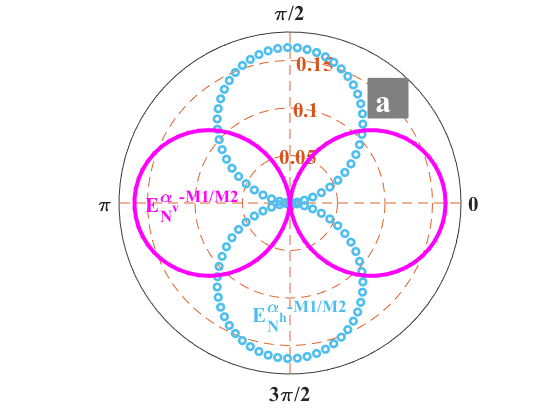}}
  \resizebox{0.45\textwidth}{!}{
  \includegraphics{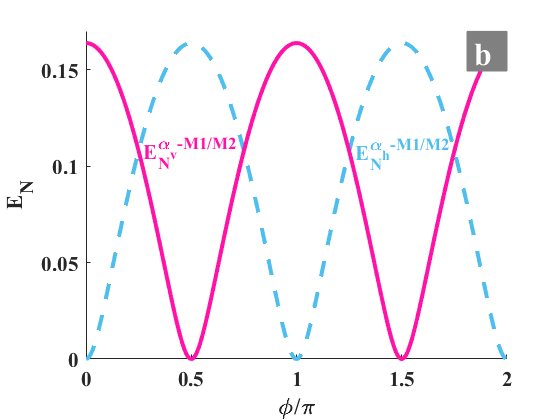}}
  \end{center}
\caption{(a) Polar entanglement representation and its (b) $2D$-representation versus $\phi$ for $J_m=0$. The quantity $\rm{E_N^{\alpha_v-M1/M2}}$ ($\rm{E_N^{\alpha_h-M1/M2}}$) is the bipartite entanglement between the $\rm{TE}$ ($\rm{TM}$) mode and each mechanical resonator.  In these figures, $G_m=0.2\omega_m$, $n_{th}^j=100$, and the other parameters are the same as in \autoref{fig:fig3}.}
\label{fig:fig4}
\end{figure}

\begin{figure*}[tbh]
\begin{center}
  \resizebox{0.45\textwidth}{!}{
  \includegraphics{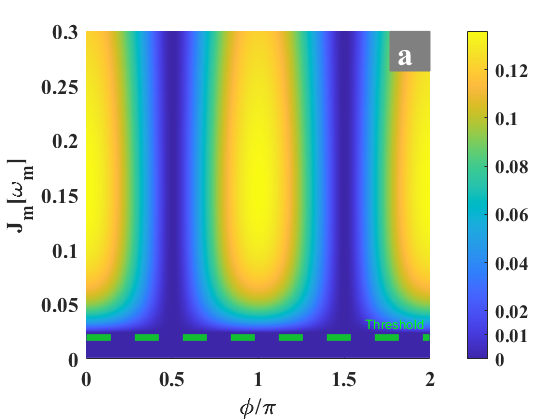}}
  \resizebox{0.45\textwidth}{!}{
  \includegraphics{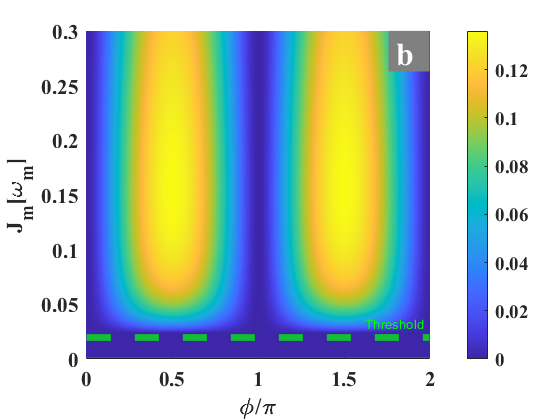}}
  \resizebox{0.45\textwidth}{!}{
  \includegraphics{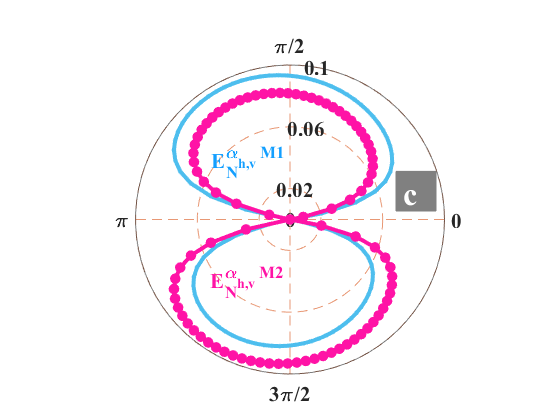}}
  \resizebox{0.45\textwidth}{!}{
  \includegraphics{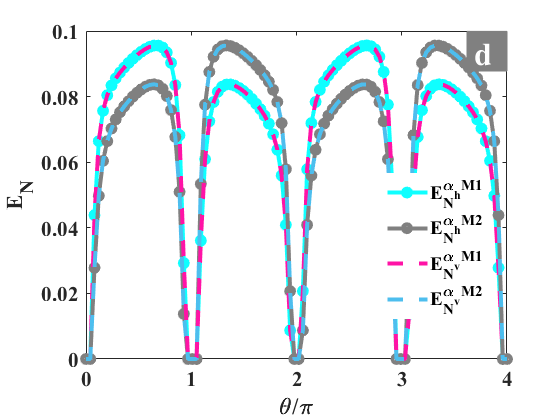}}
  \end{center}
\caption{(a)  Contour plot of the entanglement between $\rm{TE}$ mode and each mechanical resonator ($E_N^{\rm{\alpha_v-M1/M2}}$) versus $J_m$ and $\phi$ for $\theta=\frac{\pi}{2}$. (b)  Contour plot of the entanglement between $\rm{TM}$ mode and each mechanical resonators ($E_N^{\rm{\alpha_h-M1/M2}}$) versus $J_m$ and $\phi$ for $\theta=\frac{\pi}{2}$. (c) Polar entanglement representation versus $\theta$ for $J_m=0.2\omega_m$  and $\phi=\frac{\pi}{4}$. (d) $2D$-representation of (c) where $E_N^{\rm{\alpha_{h,v}-M1}}$ ($E_N^{\rm{\alpha_{h,v}-M2}}$) reaches it optimal value for $\theta=(n+1/2)\pi$, $n$ being an even integer ($\theta=(n+1/2)\pi$, $n$ being an odd integer).  For all these figures, $G_m=0.2\omega_m$, $n_{th}^j=100$, and the other parameters are the same as in \autoref{fig:fig3}. }
\label{fig:fig5}
\end{figure*}

\section{Enhancing bipartite polarization entanglement via dark mode control} \label{sec:bipart}
In this section, we aim to  enhance bipartite entanglement in our system through dark mode control. For this purpose, we consider our linearized Hamiltonian,
\begin{align}\label{eq:lin2r}
 H^r_{lin}&=-\sum_{p=\updownarrow,\leftrightarrow} (\tilde{\Delta}_p+i\frac{\kappa}{2}) \delta \alpha_p^\dagger \delta \alpha_p + \sum_{j=1,2} (\omega_j-i\frac{\gamma_m}{2}) \delta \beta_j^\dagger \delta \beta_j \nonumber \\&+ \sum_{j=1,2} \sum_{p=\updownarrow,\leftrightarrow} G_p (\delta\alpha_p^\dagger \delta\beta_j + \delta\alpha_p \delta\beta_j^\dagger)\nonumber \\&+J_m \left(e^{i\theta}\delta \beta_1^\dagger \beta_2 + e^{-i\theta} \beta_1 \beta_2^\dagger \right).
\end{align}
By assuming no coupling between the mechanical resonators ($J_m=0$), and by defined the following bosonic bright and dark modes,
\begin{equation}
\begin{cases}
\delta\alpha_{\leftrightarrow}&=\frac{G_{\leftrightarrow} B_+ + G_{\updownarrow} B_-}{G},\\
\delta\alpha_{\updownarrow}&=\frac{G_{\updownarrow} B_+ - G_{\leftrightarrow} B_-}{G},
\end{cases}
\end{equation}
where $G=\sqrt{G_{\updownarrow}^2+G_{\leftrightarrow}^2}$, this linearized Hamiltonian can be re-written in terms of bright and dark modes as,
\begin{align}\label{eq:lin2}
  H^r_{lin}&=-\sum_{k=\pm} \omega_k B^\dagger_kB_k -G_+\sum_{j=1,2} (\delta\beta_jB^\dagger_+ +\delta\beta^\dagger_jB_+) \nonumber \\& + \sum_{j=1,2}(\omega_j -i\frac{\gamma_m}{2}) \delta\beta^\dagger_j \delta\beta_j + G_-(B^\dagger_+B_- + B^\dagger_-B_+),
\end{align}
with $G_+=G$,  $G_-=\frac{G_{\updownarrow}G_{\leftrightarrow}}{|G|^2}(\tilde{\Delta}_{\leftrightarrow}-\tilde{\Delta}_{\updownarrow})$, and $\omega_\pm=\frac{(\tilde{\Delta}_{\updownarrow}+i\frac{\kappa}{2}) |G_{{\leftrightarrow}(\updownarrow)}|^2+(\tilde{\Delta}_{\leftrightarrow}+i\frac{\kappa}{2}) |G_{\updownarrow(\leftrightarrow)}|^2}{|G|^2}$. From \autoref{eq:lin2} and the expressions of $G_{\pm}$, it can be seen that $G_-=0$ for $\tilde{\Delta}_{\leftrightarrow}=\tilde{\Delta}_{\updownarrow}$, and that leads to the fact that the mode $B_-$ is decoupled from the 
system, i.e., it is a dark mode. Moreover, this dark mode can be also displayed by controlling the polarization angle $\phi$. Indeed, for $\tilde{\Delta}_{\leftrightarrow}\neq\tilde{\Delta}_{\updownarrow}$, dark mode shows up for either $G_{\updownarrow}=0$ (for $\phi=\frac{\pi}{2}n$ with $n$ an odd integer) or $G_{\leftrightarrow}=0$ (for $\phi=n\pi$ with $n$ an integer). Beside of this dark mode, one has the bright mode $B_+$ which is always coupled to the mechanical mode owing to $G_+>0$ as long as $G_m\neq0$. In \autoref{fig:fig3}, we have displayed the coupling $G_-$ in $\phi-$polar representation. As predicted, it can be seen that dark mode exists for $\phi=\frac{\pi}{2}n$ with $n$ an integer, even though $\Delta_{\updownarrow}\neq\Delta_{\leftrightarrow}$. Therefore, the Optical Dark Mode Unbreaking ($\rm{ODMU}$) regime corresponds to $\phi=\frac{\pi}{2}n$ with $n$ an integer, while  Optical Dark Mode breaking ($\rm{ODMB}$) happens when $\phi\neq\frac{\pi}{2}n$ with $n$ an integer. Such dark mode control based on the polarization angle, while the cavities are non-degenerated, is one of the key points of our investigation, because it reflects the physical case where engineering degenerated cavities is a difficult task. When the phonon hopping rate is non zero ($J_m\neq0$), each polarized field couples with the two coupled mechanical resonators. In this case, our system can be reduced to the one investigated in \cite{Lai.2022} where dark mode is controlled via the modulation phase of $J_m$. However, we are mostly interested on the dark mode control via the polarization angle in our investigation here, since it is the one that induces multi-path entanglement. 

\begin{figure*}[tbh]
\begin{center}
  \resizebox{0.45\textwidth}{!}{
  \includegraphics{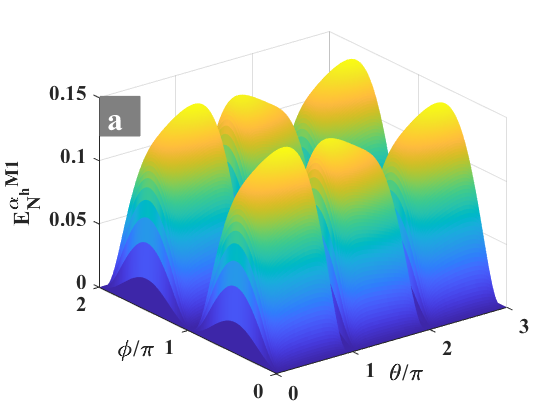}}
  \resizebox{0.45\textwidth}{!}{
  \includegraphics{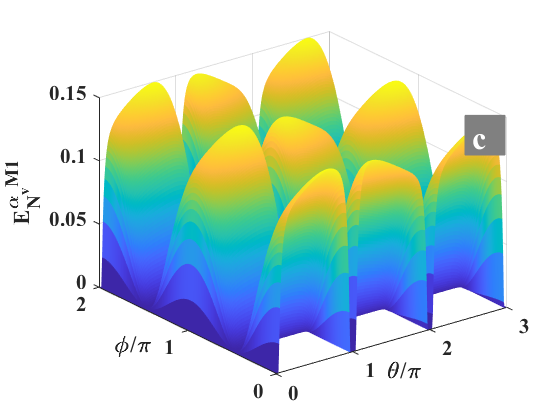}}
  \resizebox{0.45\textwidth}{!}{
  \includegraphics{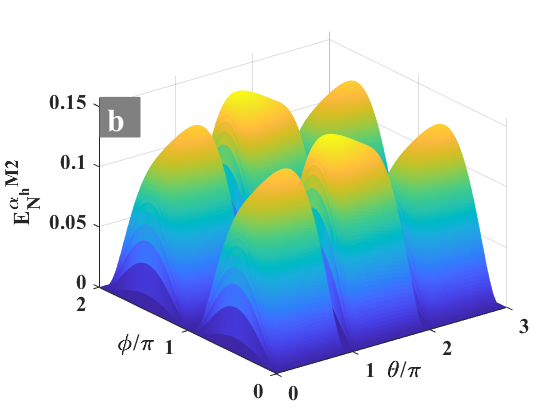}}
  \resizebox{0.45\textwidth}{!}{
  \includegraphics{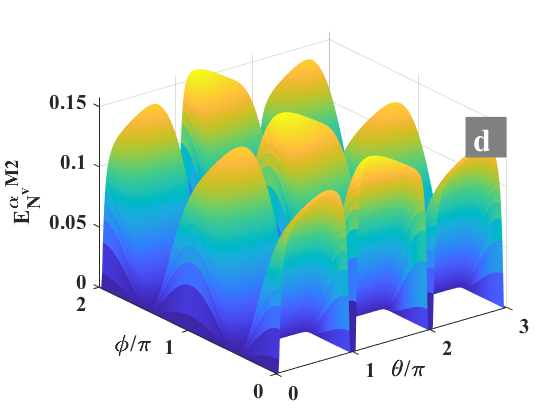}}
  \end{center}
\caption{$3D$-representation of the entanglements versus the polarization angle $\phi$ and the modulation phase $\theta$. (a,b) Entanglement between the $\rm{TM}$ polarized mode with the first ($E_N^{\rm{\alpha_{h}-M1}}$) and second ($E_N^{\rm{\alpha_{h}-M2}}$) mechanical resonator, respectively. (c,d) Entanglement involving the $\rm{TE}$ polarized mode with the first ($E_N^{\rm{\alpha_{v}-M1}}$) and second ($E_N^{\rm{\alpha_{v}-M2}}$) mechanical resonator. For these figures, we have set $n_{th}^j=100$, $J_m=0.2\omega_m$ and $G_m=0.2\omega_m$. The other parameters are the same as in \autoref{fig:fig3}.}
\label{fig:fig6}
\end{figure*}

The investigated bipartite entanglement is displayed in \autoref{fig:fig4}, for $J_m=0$.  \autoref{fig:fig4}a depicts the polar representation of the entanglement between the $\rm{TE}$ ($\rm{TM}$) mode and one of the mechanical resonators, i.e., $\rm{E_N^{\alpha_v-M1/M2}}$ ($\rm{E_N^{\alpha_h-M1/M2}}$). It can be seen that the entanglement  $\rm{E_N^{\alpha_v-M1/M2}}$ (solid line in \autoref{fig:fig4}a) reach its optimal value for $\phi=n\pi$ with $n$ being an integer. One also observes that this entanglement does not exist ($\rm{E_N^{\alpha_v-M1/M2}}=0$) for $\phi=m\frac{\pi}{2}$ with $m$ being an odd integer. Reversely, the entanglement $\rm{E_N^{\alpha_h-M1/M2}}$ (dotted line in \autoref{fig:fig4}a) has its optimal values at $\phi=m\frac{\pi}{2}$ with $m$ being an odd integer, while it vanishes ($\rm{E_N^{\alpha_h-M1/M2}}=0$) for $\phi=n\pi$ with $n$ being an integer. This behavior of $\rm{E_N^{\alpha_v-M1/M2}}$ ($\rm{E_N^{\alpha_h-M1/M2}}$) is related to the polarization feature used, since the $\rm{TE}$ ($\rm{TM}$)  mode does not exists for $\phi=n\pi$ ($\phi=m\frac{\pi}{2}$) as aforementioned. Moreover, it can be seen that the same amount of these entanglement simultaneously exist at $\phi=(p+1)\frac{\pi}{4}$ for $p$ and even integer, resulting to fact that four entanglement paths are engineered in our system simultaneously. Such feature of multi-path entanglement in our proposal is triggered by the polarization effect, which constitutes the merit of introducing polarized fields in our investigation. These features of $\rm{E_N^{\alpha_v-M1/M2}}$ and $\rm{E_N^{\alpha_h-M1/M2}}$ can be also displayed in $2-D$-representation as depicted in  \autoref{fig:fig4}b, where these same quantities are shown versus the polarized angle $\phi$. One can observe that the generated entanglements exhibit oscillatory behaviors, which are reminiscent of entanglement death and revival phenomena. From the above observation, it results that only $\rm{TE}$ ($\rm{E_N^{\alpha_v-M1/M2}}$) entanglement or $\rm{TM}$ ($\rm{E_N^{\alpha_h-M1/M2}}$) entanglement exists in the $\rm{ODMU}$  regime. In order to simultaneously generate these entanglement, the dark mode must be broken by adjusting  $\phi\neq\frac{\pi}{2}n$ leading to the $\rm{ODMB}$  regime where four entanglement quantities are engineered. This multi-path entanglement engineering is fully based on the polarization angle $\phi$ together with dark mode control. 

\begin{figure}[tbh]
\begin{center}
  \resizebox{0.45\textwidth}{!}{
  \includegraphics{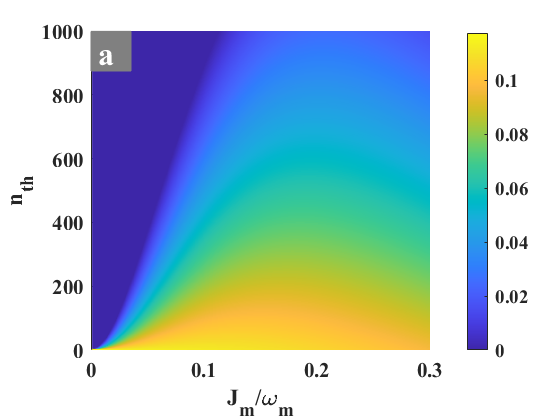}}
  \resizebox{0.45\textwidth}{!}{
  \includegraphics{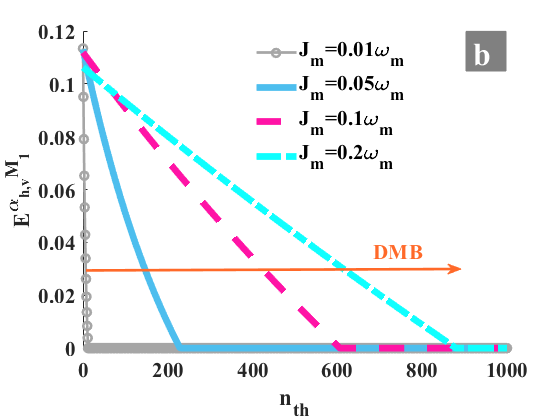}}
  \end{center}
\caption{(a) Contour plot of the entanglement $E_N^{\rm{\alpha_{h,v}M1}}$ versus $J_m$ and the thermal phonon number $n_{th}$ for $G_m=0.2\omega_m$,  $\phi=\frac{\pi}{4}$, and $\theta=\frac{\pi}{2}$. (b) Corresponding \rm{2D}-representation of (a) with different values of $J_m$, under the DMB regime condition. The other parameters are as those in  \autoref{fig:fig3}.}
\label{fig:fig7}
\end{figure}

The bipartite entanglement feature in our proposal can be further impacted as the mechanical coupling came into play ($J_m\neq 0$). In \autoref{fig:fig5}, the entanglement is displayed when the dark mode effect is broken, i.e., $J_m\neq 0$ and $\theta=\pi/2$. For instance, \autoref{fig:fig5}a (\autoref{fig:fig5}b) depicts the contour plot of $\rm{E_N^{\alpha_v-M1/M2}}$ ($\rm{E_N^{\alpha_h-M1/M2}}$) versus the coupling $J_m$ and the polarization angle $\phi$ for $\theta=\pi/2$. It can be seen that the $\rm{TE}$ ($\rm{TM}$) mode is optimally entangled to the mechanical resonators for $\phi=n\pi$ with $n$ being an integer ($\phi=n\frac{\pi}{2}$ with $n$ being an odd integer). Moreover, these entanglements are generated from a certain threshold of the phonon hopping rate, i.e., $J_m\gtrsim0.05\omega_m$. To point out the entanglement feature over the phase-dependent mechanical coupling, we plotted the polar representation of the entanglement versus $\theta$ in \autoref{fig:fig5}c for $\phi=\pi/4$ and $J_m=0.2\omega_m$. It can be seen that the entanglement between each polarized mode and the first mechanical resonator $\rm{E_N^{\alpha_{h,v}-M_1}}$ (second mechanical resonator $\rm{E_N^{\alpha_{h,v}-M_2}}$)  reaches its optimal value for $\phi=(2n+1)\frac{\pi}{2}$ with $n$ being an even integer  ($\phi=(2n+1)\frac{\pi}{2}$ with $n$ being an odd integer). Moreover, it is interesting to mention that there is an switching between these entanglements depending on whether we are on the upper half or the lower half of the unit circle. This switching can be further observed in \autoref{fig:fig5}d, which displays entanglement versus $\theta$ with the same parameters as in \autoref{fig:fig5}c.  

In order to investigate both effect of $\phi$ and $\theta$ on the bipartite entanglement, we displayed on \autoref{fig:fig6} a plot of the entanglements versus $\phi$ and $\theta$. Indeed, \autoref{fig:fig6}(a-d) represent the entanglement between the $\rm{TM}$ mode with first mechanical resonator ($\rm{E_N^{\alpha_{h}-M_1}}$), the $\rm{TE}$ mode with first mechanical resonator ($\rm{E_N^{\alpha_{v}-M_1}}$), the $\rm{TM}$ mode with second mechanical resonator ($\rm{E_N^{\alpha_{h}-M_2}}$), and the $\rm{TE}$ mode with second mechanical resonator ($\rm{E_N^{\alpha_{h}-M_1}}$), respectively. It can be observed that under the dark mode breaking condition ($\theta=n\frac{\pi}{2}$ with $n$ being an odd integer) the entanglements involving $\rm{TM}$ mode reached their optimal values for $\phi=(2n+1)\frac{\pi}{2}$ with $n$ being an even integer, while those involving the  $\rm{TE}$ mode are maximum for $\phi=n\pi$ with $n$ being an integer. Moreover, the peaks of these entanglements are not equivalently distributed depending on when the mode $\rm{TM}$ or $\rm{TE}$ is coupled to the first or second resonator. For instance, the peaks appearing on \autoref{fig:fig6}a  and  \autoref{fig:fig6}b are not similarly distributed even tough both entanglements involve $\rm{TM}$. This also applies to \autoref{fig:fig6}c  and  \autoref{fig:fig6}d for $\rm{TE}$ mode. This reveals that the interaction between the polarized modes and the mechanical resonators does not necessary leads to symmetric entanglements. This asymmetric of the entanglement further highlight the multi-entanglement path feature in our system. Thermal management of these generated bipartite entanglement is carried out in \autoref{fig:fig7}. A contour plot of the entanglement  $E_N^{\rm{\alpha_{h,v}M1}}$ versus  the coupling $J_m$ and the thermal phonon number $n_{th}$ is displayed in \autoref{fig:fig7}a, for $\theta=\pi/2$. It can be seen that the entanglement survives longer against thermal noise as the coupling $J_m$ increases. For weak strength of the mechanical coupling ($J_m<0.01\omega_m$), the generated entanglement is so fragile and disappears for few thermal phonons ($n_{th} \sim 10$). Therefore, it results that the breaking of the dark mode significantly improves the resilience of the entanglement against thermal fluctuations. This can be further seen in \autoref{fig:fig7}b, where almost two order of magnitude of robustness against thermal noise is reached under dark mode control. Such a thermal management of quantum entanglement is useful for engineering noise-tolerant quantum correlations which are vital resources for modern quantum applications.        

\begin{figure}[tbh]
\begin{center}
  \resizebox{0.45\textwidth}{!}{
  \includegraphics{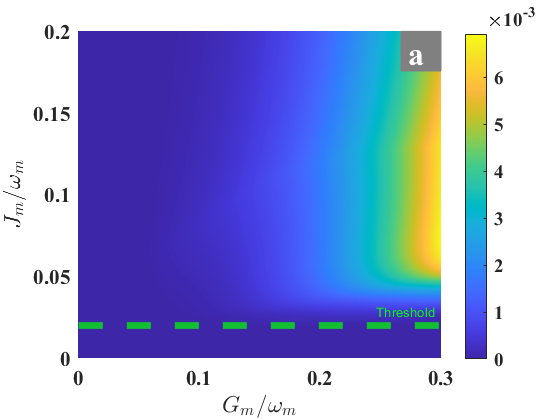}}
  \resizebox{0.45\textwidth}{!}{
  \includegraphics{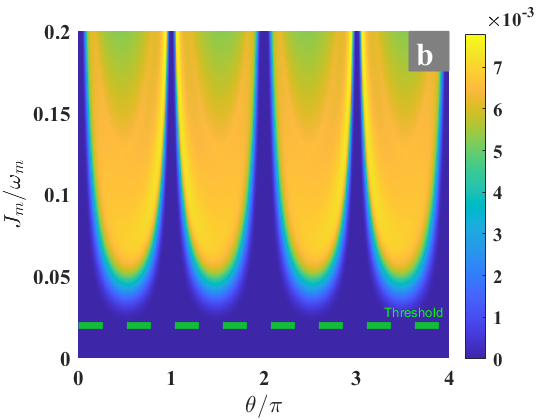}}
  \end{center}
\caption{(a) Contour plot of the residual cotangle $R^{min}_{\alpha_h}$ (same for  $R^{min}_{\alpha_v}$) versus $J_m$ and $G_m$ for $\phi=\pi/4$, $\theta=\frac{\pi}{2}$. (b) Contour plot of the residual cotangle $R^{min}_{\alpha_h}$ (same for  $R^{min}_{\alpha_v}$) versus $J_m$ and $\theta$ for $\phi=\pi/4$, $G_m=0.3\omega_m$, and $n_{th}^j=100$. The green dashed line indicates the threshold value of $J_m\sim0.02\omega_m$ from which tripartite entanglement is generated. The other parameters are as those in  \autoref{fig:fig3}.}
\label{fig:fig8}
\end{figure}

\section{Engineering tripartite polarization entanglement via dark mode control} \label{sec:tripart}

In this section, we investigate tripartite entanglement in our system. For this purpose, we seek to compute the minimum residual contangle between each polarized mode and the two mechanical resonators, as a mean of quantifying tripartite entanglement. For simplicity in our notations, we define $R^{min}_{\alpha_h}$ ($R^{min}_{\alpha_v}$) as the minimum residual contangle between $\rm{TM}$ ($\rm{TE}$) mode with the two mechanical resonators. As we intend to generate multi-path tripartite entanglement, the polarization angle $\phi$ and the the breaking of the dark mode through $\theta$ (or $J_m$) will be our parameters of interests. Moreover, we have checked and found that tripartite entanglement involving the two polarized fields and any of the mechanical resonators does not exist or is irrelevant (not shown here). 

\begin{figure}[tbh]
\begin{center}
  \resizebox{0.45\textwidth}{!}{
  \includegraphics{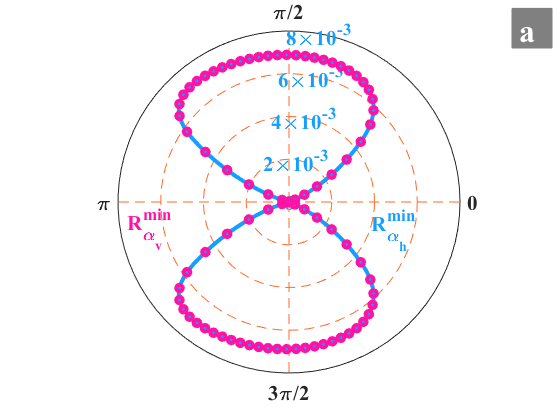}}
  \resizebox{0.45\textwidth}{!}{
  \includegraphics{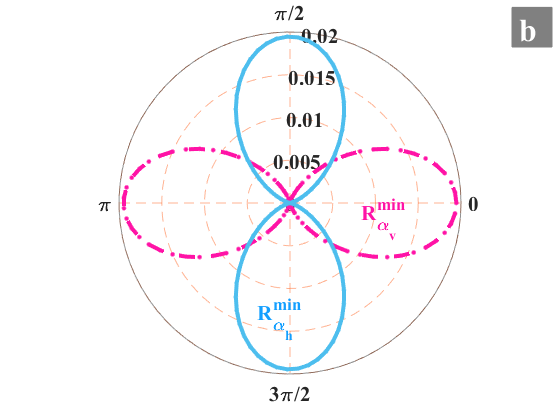}}
  \end{center}
\caption{(a) Polar representation of the residual cotangle $R^{min}_{\alpha_h}$ (same for  $R^{min}_{\alpha_v}$) versus $\theta$ for $\phi=\pi/4$. (b) Polar representation of the residual cotangle $R^{min}_{\alpha_h}$ (same for  $R^{min}_{\alpha_v}$) versus $\phi$ for $\theta=\pi/2$. For these plots, we have set $J_m=0.08\omega_m$, $G_m=0.3\omega_m$ and $n_{th}^j=100$. The other parameters are as those in  \autoref{fig:fig3}.}
\label{fig:fig9}
\end{figure}

\begin{figure}[tbh]
\begin{center}
  \resizebox{0.45\textwidth}{!}{
  \includegraphics{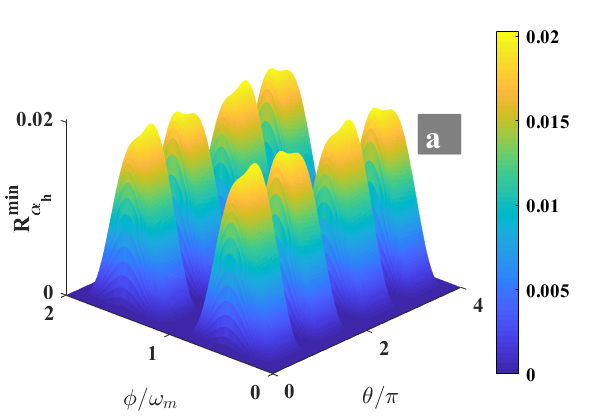}}
  \resizebox{0.45\textwidth}{!}{
  \includegraphics{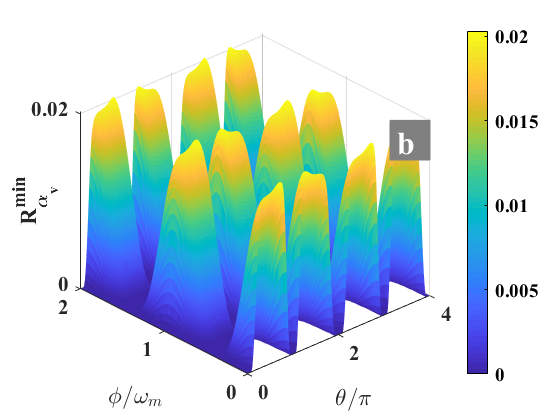}}
  \end{center}
\caption{(a) Residual cotangle $R^{min}_{\alpha_h}$ and (b) $R^{min}_{\alpha_v}$  versus  $\phi$ and $\theta$ for $J_m=0.08\omega_m$, $G_m=0.3\omega_m$ and $n_{th}^j=100$. The other parameters are as those in  \autoref{fig:fig3}.}
\label{fig:fig10}
\end{figure}

In \autoref{fig:fig8}, we represent contour plot of residual cotangle versus $J_m$ and $G_m$ (\autoref{fig:fig8} a) and versus $J_m$ and $\theta$ (\autoref{fig:fig8} b). As expected, tripartite entanglement is generated in our proposal only when $J_m\neq0$ and this is shown through the green threshold line in \autoref{fig:fig8}a ($J_m\sim 0.022\omega_m$). Moreover, it can be also observed that the generated tripartite entanglement requires also a threshold for the driving coupling strength $G_m\sim 0.15\omega_m$. Furthermore, this entanglement is generated when $\theta \neq n\pi$ with $n$ being an integer (\autoref{fig:fig8} b). Therefore, it results that tripartite entanglement is generated in our system under the dark mode breaking condition. It is worth mentioning that the polarization angle is set to $\phi=\pi/4$ in \autoref{fig:fig8}, leading to the fact that both $R^{min}_{\alpha_h}$ and $R^{min}_{\alpha_v}$ are similar.  To further get insight about the effect of both $\phi$ and $\theta$ on the generated residual cotangle, we displayed tripartite entanglement in polar representation  in \autoref{fig:fig9}. The tripartite entanglement is depicted versus $\theta$ ($\phi$) in \autoref{fig:fig9}a (\autoref{fig:fig9}b). It can be seen that entanglement is generated only under the dark mode breaking condition ($\theta \neq n\pi$), and reaches its maximum value around $\theta = m\pi/2$,  with $m$ being an odd integer. Moreover, the quantities $R^{min}_{\alpha_h}$ and $R^{min}_{\alpha_v}$ are similar as aforementioned (for $\phi=\pi/4$). This is displayed in \autoref{fig:fig9}b, which captures a polar representation of the entanglement versus $\phi$. It can be observed that both entanglements $R^{min}_{\alpha_h}$ and $R^{min}_{\alpha_v}$ perfectly hybridize at $\phi=\pi/4$, while $R^{min}_{\alpha_h}$ ($R^{min}_{\alpha_v}$) is optimal for $\phi = m\pi/2$,  with $m$ being an odd integer ($\phi = n\pi$,  with $n$ being any integer).  From \autoref{fig:fig8} and \autoref{fig:fig9}, it results that tripartite entanglement is generated in our system only in the dark mode broken regime ($J_m\neq0$, $\theta\neq n\pi$). Furthermore, multi-path entanglement triggers in the system under the condition $\phi\neq p\pi/2$, $p$ being any integer. Such multi-path tripartite entanglement generation can be useful for quantum information processing and diverse quantum computational tasks. This multi-path entanglement engineering can be further fostered by displaying entanglement versus $\theta$  and $\phi$ as shown in \autoref{fig:fig10}. Indeed, as it can be seen in  \autoref{fig:fig10}a and \autoref{fig:fig10}b, different patterns of tripartite entanglements are displayed depending on the value of both $\theta$  and $\phi$. The residual cotangle involving the  $\rm{TM}$ ($\rm{TE}$) mode with the two mechanical resonators results to different patterns where the entanglement peaks show up either at $\phi=p\pi/2$, $p$ being an odd integer for $R^{min}_{\alpha_h}$ (see \autoref{fig:fig10}a), or at $\phi=m\pi$, $m$ being any integer  for $R^{min}_{\alpha_v}$ (see \autoref{fig:fig10}b). Therefore, the polarization angle $\phi$ is revealed as a key parameter that induces multi-paths entanglement features in our system, which provides a way to access abundant tripartite quantum resources to boost number of quantum applications including quantum information processing, quantum computation and other quantum tasks. 

\begin{figure}[tbh]
\begin{center}
  \resizebox{0.45\textwidth}{!}{
  \includegraphics{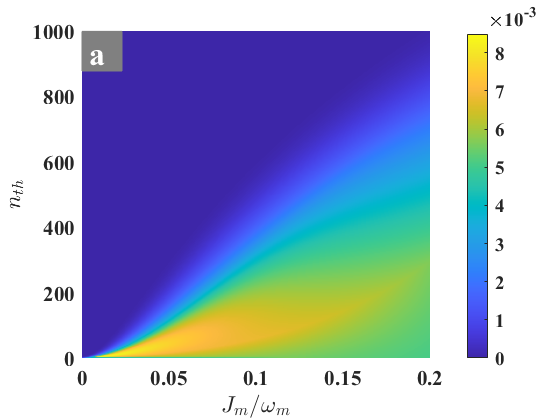}}
  \resizebox{0.45\textwidth}{!}{
  \includegraphics{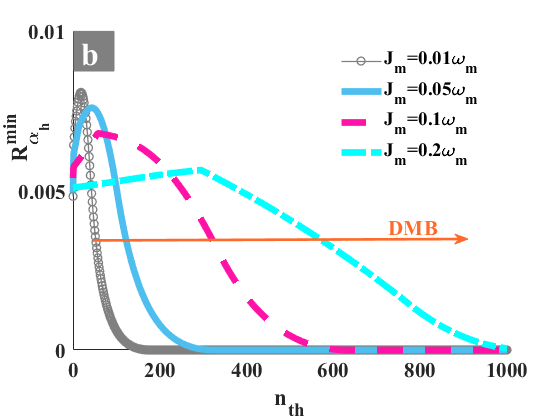}}
  \end{center}
\caption{(a) Contour plot of the residual cotangle $R^{min}_{\alpha_h}$ (same for  $R^{min}_{\alpha_v}$) versus thermal phonon number $n_{th}$ and $J_m$ for $\phi=\pi/4$, $\theta=\frac{\pi}{2}$ and $G_m=0.3\omega_m$. The other parameters are as those in  \autoref{fig:fig3}.}
\label{fig:fig11}
\end{figure}

Another interesting property of the residual cotangle to figure out is its robutness against thermal fluctuations. This feature reveals how strong the generated entanglement can resist to thermal bath, and this is interesting for number of open systems which interact with their environment. \autoref{fig:fig11} displays contour plot of residual cotangle versus the thermal phonon number ($n_{th}$) and the phonon hopping rate $J_m$. It can be seen that the tripartite entanglement resists more to thermal fluctuation as the coupling $J_m$ is increases (see \autoref{fig:fig11}a). This further highlights the merit of having $J_m\neq0$ in our system, because $J_m=0$ does not lead to any entanglement in our proposal. Moreover, \autoref{fig:fig11} depicts how the entanglement is triggered in the dark mode breaking regime, as indicated through direction of the arrow, and this entanglement resists to a thermal phonon population of $n_{th}\sim 1000$, which is up to two order of magnitude stronger than in the Unbreaking regime. It is noteworthy to mention that we have set $\phi=n\pi/4$ in \autoref{fig:fig11}, so that the plots capture both $R^{min}_{\alpha_h}$ and $R^{min}_{\alpha_v}$ which are equivalent. Such a noise-tolerant entanglement is suitable for number of modern quantum applications.

\section{Conclusion}\label{sec:Concl}
We have proposed a scheme to engineer multi-path entanglements in optomechanical system through dark mode control. Our model consists of vector electromgnetic fields, i.e., the vertical (transverse electric (TE) and  horizontal (transverse magnetic [(TM]) modes, which drive two mechanically coupled mechanical resonators. The phase modulation of the mechanical coupling ($\theta$) together with the polarization angle ($\phi$) are used to switch from the dark mode unbroken to the dark mode broken regimes in our proposal. When the mechanical coupling is turned off ($J_m=0$), only bipartite multi-path entanglements are generated under dark mode control through the polarization angle. For $J_m\neq0$ the dark mode breaking depends on simultaneous tuning of both  $\theta$ and $\phi$, leading to an engineering of both multi-path bipartite and tripartite entanglements in the system. The engineered entanglements are triggered from a certain threshold of the mechanical coupling, and are enhanced as the strength of $J_m$ increases. For specific value of the polarization angle $\phi=\pi/4$, twin entangled states are generated, which might be useful for particular tasks and technological applications. Moreover, the generated entangled states are more resilient against thermal fluctuations in the \rm{DMB} regime, i.e., at least two order of magnitude of robustness enhancement against thermal phonon population is achieved compared to the Unbreaking regime. The multi-path entanglement generation scheme introduced in this work reveals how noise-tolerant quantum resources can be engineered in optomechanical system, and such resources can be useful for quantum information processing, quantum communication, and number of quantum computational tasks. Furthermore, the proposed scheme can be extended to other fields such as in microwave and hybrid optomechanical contexts.

\section*{Availability of data and materials}
Relevant data are included in the manuscript,and supplement data are available upon reasonable request.

\section*{Competing interests}
The authors declare that they have no competing interests.

\section*{Funding}
Not applicable.

\section*{Authors' contributions}

\textbf{P. D.} Conceptualized the work, carried out the simulations and wrote the original draft. \textbf{R. A.} Revised the draft and validated the numerical results. \textbf{A. J.A.:} Participated on the data curation and the validation of the results. \textbf{A. A.-K.:} Proposed the methodology, and revised the last version of the manuscript. \textbf{A.-H. A.-A.:} Supervised the work and participated on the revision and writing of the manuscript. All authors participated in the discussions.

\section*{Acknowledgments}

P.D. acknowledges the Iso-Lomso Fellowship from the  Stellenbosch Institute for Advanced Study (STIAS),  Stellenbosch 7600, South Africa, and The Institute for Advanced Study, Wissenschaftskolleg zu Berlin, Wallotstrasse 19, 14193 Berlin, Germany. This work was supported by Princess Nourah bint Abdulrahman University Researchers Supporting Project number (PNURSP2026R399), Princess Nourah bint Abdulrahman University, Riyadh, Saudi Arabia. The authors are thankful to the Deanship of Graduate Studies and Scientific Research at the University of Bisha for supporting this work through the Fast-Track Research Support Program.

\bibliography{Vector_polarization}

@Article{Xiong2016,
  author    = {Xiong, Hao and Huang, Ya-Min and Wan, Liang-Liang and Wu, Ying},
  journal   = {Physical Review A},
  title     = {Vector cavity optomechanics in the parameter configuration of optomechanically induced transparency},
  year      = {2016},
  issn      = {2469-9934},
  month     = jul,
  number    = {1},
  pages     = {013816},
  volume    = {94},
  doi       = {10.1103/PhysRevA.94.013816},
  publisher = {American Physical Society (APS)},
}

@Article{Wang2018,
  author    = {Wang, Bao and Xiong, Hao and Jia, Xiao and Wu, Ying},
  journal   = {Scientific Reports},
  title     = {Phonon laser in the coupled vector cavity optomechanics},
  year      = {2018},
  issn      = {2045-2322},
  month     = jan,
  number    = {1},
  pages     = {282},
  volume    = {8},
  doi       = {10.1038/s41598-017-17395-x},
  publisher = {Springer Science and Business Media LLC},
}

@Article{Buters2016,
  author    = {Buters, F. M. and Weaver, M. J. and Eerkens, H. J. and Heeck, K. and de Man, S. and Bouwmeester, D.},
  journal   = {Physical Review A},
  title     = {Optomechanics with a polarization nondegenerate cavity},
  year      = {2016},
  issn      = {2469-9934},
  month     = dec,
  number    = {6},
  pages     = {063813},
  volume    = {94},
  doi       = {10.1103/PhysRevA.94.063813},
  publisher = {American Physical Society (APS)},
}

@Article{Duggan2019,
  author    = {Duggan, Robert and del Pino, Javier and Verhagen, Ewold and Alù, Andrea},
  journal   = {Physical Review Letters},
  title     = {Optomechanically Induced Birefringence and Optomechanically Induced Faraday Effect},
  year      = {2019},
  issn      = {1079-7114},
  month     = jul,
  number    = {2},
  pages     = {023602},
  volume    = {123},
  doi       = {10.1103/PhysRevLett.123.023602},
  publisher = {American Physical Society (APS)},
}

@Article{Li2021,
  author    = {Li, Ying and Jiao, Ya-Feng and Liu, Jing-Xue and Miranowicz, Adam and Zuo, Yun-Lan and Kuang, Le-Man and Jing, Hui},
  journal   = {Nanophotonics},
  title     = {Vector optomechanical entanglement},
  year      = {2021},
  issn      = {2192-8614},
  month     = nov,
  number    = {1},
  pages     = {67--77},
  volume    = {11},
  doi       = {10.1515/nanoph-2021-0485},
  publisher = {Walter de Gruyter GmbH},
}

@Article{Rossi2009,
  author    = {Rossi, Alessandro and Vallone, Giuseppe and Chiuri, Andrea and De Martini, Francesco and Mataloni, Paolo},
  journal   = {Physical Review Letters},
  title     = {Multipath Entanglement of Two Photons},
  year      = {2009},
  issn      = {1079-7114},
  month     = apr,
  number    = {15},
  pages     = {153902},
  volume    = {102},
  doi       = {10.1103/PhysRevLett.102.153902},
  publisher = {American Physical Society (APS)},
}

@Article{Sutcliffe2023,
  author    = {Sutcliffe, Evan and Beghelli, Alejandra},
  journal   = {IEEE Transactions on Quantum Engineering},
  title     = {Multiuser Entanglement Distribution in Quantum Networks Using Multipath Routing},
  year      = {2023},
  issn      = {2689-1808},
  pages     = {1--15},
  volume    = {4},
  doi       = {10.1109/TQE.2023.3329714},
  publisher = {Institute of Electrical and Electronics Engineers (IEEE)},
}

@Article{Zhang2023,
  author    = {Zhang, Ling and Liu, Qin},
  journal   = {Quantum Information Processing},
  title     = {Concurrent multipath quantum entanglement routing based on segment routing in quantum hybrid networks},
  year      = {2023},
  issn      = {1573-1332},
  month     = mar,
  number    = {3},
  pages     = {148},
  volume    = {22},
  doi       = {10.1007/s11128-023-03891-9},
  publisher = {Springer Science and Business Media LLC},
}

@Article{Hu_2020,
  author    = {Hu, Xiao-Min and Xing, Wen-Bo and Liu, Bi-Heng and Huang, Yun-Feng and Li, Chuan-Feng and Guo, Guang-Can and Erker, Paul and Huber, Marcus},
  journal   = {Physical Review Letters},
  title     = {Efficient Generation of High-Dimensional Entanglement through Multipath Down-Conversion},
  year      = {2020},
  issn      = {1079-7114},
  month     = aug,
  number    = {9},
  pages     = {090503},
  volume    = {125},
  doi       = {10.1103/PhysRevLett.125.090503},
  publisher = {American Physical Society (APS)},
}

@Article{Shah2024,
  author    = {Shah, Parth S. and Yang, Frank and Joshi, Chaitali and Mirhosseini, Mohammad},
  journal   = {PRX Quantum},
  title     = {Stabilizing Remote Entanglement via Waveguide Dissipation},
  year      = {2024},
  issn      = {2691-3399},
  month     = sep,
  number    = {3},
  pages     = {030346},
  volume    = {5},
  doi       = {10.1103/PRXQuantum.5.030346},
  publisher = {American Physical Society (APS)},
}

@Article{Steffen2006,
  author    = {Steffen, Matthias and Ansmann, M. and Bialczak, Radoslaw C. and Katz, N. and Lucero, Erik and McDermott, R. and Neeley, Matthew and Weig, E. M. and Cleland, A. N. and Martinis, John M.},
  journal   = {Science},
  title     = {Measurement of the Entanglement of Two Superconducting Qubits via State Tomography},
  year      = {2006},
  issn      = {1095-9203},
  month     = sep,
  number    = {5792},
  pages     = {1423--1425},
  volume    = {313},
  doi       = {10.1126/science.1130886},
  publisher = {American Association for the Advancement of Science (AAAS)},
}

@Article{DiCarlo2010,
  author    = {DiCarlo, L. and Reed, M. D. and Sun, L. and Johnson, B. R. and Chow, J. M. and Gambetta, J. M. and Frunzio, L. and Girvin, S. M. and Devoret, M. H. and Schoelkopf, R. J.},
  journal   = {Nature},
  title     = {Preparation and measurement of three-qubit entanglement in a superconducting circuit},
  year      = {2010},
  issn      = {1476-4687},
  month     = sep,
  number    = {7315},
  pages     = {574--578},
  volume    = {467},
  doi       = {10.1038/nature09416},
  publisher = {Springer Science and Business Media LLC},
}

@Article{Palomaki2013,
  author    = {Palomaki, T. A. and Teufel, J. D. and Simmonds, R. W. and Lehnert, K. W.},
  journal   = {Science},
  title     = {Entangling Mechanical Motion with Microwave Fields},
  year      = {2013},
  issn      = {1095-9203},
  month     = nov,
  number    = {6159},
  pages     = {710--713},
  volume    = {342},
  doi       = {10.1126/science.1244563},
  publisher = {American Association for the Advancement of Science (AAAS)},
}

@Article{Bowen2002,
  author    = {Bowen, Warwick P. and Treps, Nicolas and Schnabel, Roman and Lam, Ping Koy},
  journal   = {Physical Review Letters},
  title     = {Experimental Demonstration of Continuous Variable Polarization Entanglement},
  year      = {2002},
  issn      = {1079-7114},
  month     = dec,
  number    = {25},
  pages     = {253601},
  volume    = {89},
  doi       = {10.1103/PhysRevLett.89.253601},
  publisher = {American Physical Society (APS)},
}

@Article{Hensen2015,
  author    = {Hensen, B. and Bernien, H. and Dréau, A. E. and Reiserer, A. and Kalb, N. and Blok, M. S. and Ruitenberg, J. and Vermeulen, R. F. L. and Schouten, R. N. and Abellán, C. and Amaya, W. and Pruneri, V. and Mitchell, M. W. and Markham, M. and Twitchen, D. J. and Elkouss, D. and Wehner, S. and Taminiau, T. H. and Hanson, R.},
  journal   = {Nature},
  title     = {Loophole-free Bell inequality violation using electron spins separated by 1.3 kilometres},
  year      = {2015},
  issn      = {1476-4687},
  month     = oct,
  number    = {7575},
  pages     = {682--686},
  volume    = {526},
  doi       = {10.1038/nature15759},
  publisher = {Springer Science and Business Media LLC},
}

@Article{Korppi2018,
  author    = {Ockeloen-Korppi, C. F. and Damskägg, E. and Pirkkalainen, J.-M. and Asjad, M. and Clerk, A. A. and Massel, F. and Woolley, M. J. and Sillanpää, M. A.},
  journal   = {Nature},
  title     = {Stabilized entanglement of massive mechanical oscillators},
  year      = {2018},
  issn      = {1476-4687},
  month     = apr,
  number    = {7702},
  pages     = {478--482},
  volume    = {556},
  doi       = {10.1038/s41586-018-0038-x},
  publisher = {Springer Science and Business Media LLC},
}

@Article{Kotler2021,
  author    = {Kotler, Shlomi and Peterson, Gabriel A. and Shojaee, Ezad and Lecocq, Florent and Cicak, Katarina and Kwiatkowski, Alex and Geller, Shawn and Glancy, Scott and Knill, Emanuel and Simmonds, Raymond W. and Aumentado, José and Teufel, John D.},
  journal   = {Science},
  title     = {Direct observation of deterministic macroscopic entanglement},
  year      = {2021},
  issn      = {1095-9203},
  month     = may,
  number    = {6542},
  pages     = {622--625},
  volume    = {372},
  doi       = {10.1126/science.abf2998},
  publisher = {American Association for the Advancement of Science (AAAS)},
}

@Article{Mercier2021,
  author    = {Mercier de Lépinay, Laure and Ockeloen-Korppi, Caspar F. and Woolley, Matthew J. and Sillanpää, Mika A.},
  journal   = {Science},
  title     = {Quantum mechanics–free subsystem with mechanical oscillators},
  year      = {2021},
  issn      = {1095-9203},
  month     = may,
  number    = {6542},
  pages     = {625--629},
  volume    = {372},
  doi       = {10.1126/science.abf5389},
  publisher = {American Association for the Advancement of Science (AAAS)},
}

@Article{Man2023,
  author    = {Mannalath, Vaisakh and Pathak, Anirban},
  journal   = {Physical Review A},
  title     = {Multiparty entanglement routing in quantum networks},
  year      = {2023},
  issn      = {2469-9934},
  month     = dec,
  number    = {6},
  pages     = {062614},
  volume    = {108},
  doi       = {10.1103/PhysRevA.108.062614},
  publisher = {American Physical Society (APS)},
}

@Article{Oh2024,
  author    = {Oh, Donghak and Baek, Soojeong and Lee, Sangha and Lee, Kyungmin and Park, Jagang and Liu, Zhaowei and Kim, Teun-Teun and Min, Bumki},
  journal   = {Nanophotonics},
  title     = {Complete asymmetric polarization conversion at zero-eigenvalue exceptional points of non-Hermitian metasurfaces},
  year      = {2024},
  issn      = {2192-8614},
  month     = oct,
  number    = {24},
  pages     = {4409--4416},
  volume    = {13},
  doi       = {10.1515/nanoph-2024-0391},
  publisher = {Walter de Gruyter GmbH},
}

@Article{Slussarenko2019,
  author    = {Slussarenko, Sergei and Pryde, Geoff J.},
  journal   = {Applied Physics Reviews},
  title     = {Photonic quantum information processing: A concise review},
  year      = {2019},
  issn      = {1931-9401},
  month     = oct,
  number    = {4},
  pages     = {041303},
  volume    = {6},
  doi       = {10.1063/1.5115814},
  publisher = {AIP Publishing},
}

@Article{Wendin2017,
  author    = {Wendin, G},
  journal   = {Reports on Progress in Physics},
  title     = {Quantum information processing with superconducting circuits: a review},
  year      = {2017},
  issn      = {1361-6633},
  month     = sep,
  number    = {10},
  pages     = {106001},
  volume    = {80},
  doi       = {10.1088/1361-6633/aa7e1a},
  publisher = {IOP Publishing},
}

@Article{Pezz2018,
  author    = {Pezzè, Luca and Smerzi, Augusto and Oberthaler, Markus K. and Schmied, Roman and Treutlein, Philipp},
  journal   = {Reviews of Modern Physics},
  title     = {Quantum metrology with nonclassical states of atomic ensembles},
  year      = {2018},
  issn      = {1539-0756},
  month     = sep,
  number    = {3},
  pages     = {035005},
  volume    = {90},
  doi       = {10.1103/RevModPhys.90.035005},
  publisher = {American Physical Society (APS)},
}

@Article{Polino2020,
  author    = {Polino, Emanuele and Valeri, Mauro and Spagnolo, Nicolò and Sciarrino, Fabio},
  journal   = {AVS Quantum Science},
  title     = {Photonic quantum metrology},
  year      = {2020},
  issn      = {2639-0213},
  month     = jun,
  number    = {2},
  pages     = {024703},
  volume    = {2},
  doi       = {10.1116/5.0007577},
  publisher = {American Vacuum Society},
}

@Article{Degen2017,
  author    = {Degen, C. L. and Reinhard, F. and Cappellaro, P.},
  journal   = {Reviews of Modern Physics},
  title     = {Quantum sensing},
  year      = {2017},
  issn      = {1539-0756},
  month     = jul,
  number    = {3},
  pages     = {035002},
  volume    = {89},
  doi       = {10.1103/RevModPhys.89.035002},
  publisher = {American Physical Society (APS)},
}

@Article{Djorwe2019,
  author    = {Djorwe, P. and Pennec, Y. and Djafari-Rouhani, B.},
  journal   = {Physical Review Applied},
  title     = {Exceptional Point Enhances Sensitivity of Optomechanical Mass Sensors},
  year      = {2019},
  issn      = {2331-7019},
  month     = aug,
  number    = {2},
  pages     = {024002},
  volume    = {12},
  doi       = {10.1103/PhysRevApplied.12.024002},
  publisher = {American Physical Society (APS)},
}

@Article{Djorw2024,
  author    = {Djorwé, P. and Asjad, M. and Pennec, Y. and Dutykh, D. and Djafari-Rouhani, B.},
  journal   = {Physical Review Research},
  title     = {Parametrically enhancing sensor sensitivity at an exceptional point},
  year      = {2024},
  issn      = {2643-1564},
  month     = sep,
  number    = {3},
  pages     = {033284},
  volume    = {6},
  doi       = {10.1103/PhysRevResearch.6.033284},
  publisher = {American Physical Society (APS)},
}

@Article{Stannigel2010,
  author    = {Stannigel, K. and Rabl, P. and Sørensen, A. S. and Zoller, P. and Lukin, M. D.},
  journal   = {Physical Review Letters},
  title     = {Optomechanical Transducers for Long-Distance Quantum Communication},
  year      = {2010},
  issn      = {1079-7114},
  month     = nov,
  number    = {22},
  pages     = {220501},
  volume    = {105},
  doi       = {10.1103/PhysRevLett.105.220501},
  publisher = {American Physical Society (APS)},
}

@Article{Knaut2024,
  author    = {Knaut, C. M. and Suleymanzade, A. and Wei, Y.-C. and Assumpcao, D. R. and Stas, P.-J. and Huan, Y. Q. and Machielse, B. and Knall, E. N. and Sutula, M. and Baranes, G. and Sinclair, N. and De-Eknamkul, C. and Levonian, D. S. and Bhaskar, M. K. and Park, H. and Lončar, M. and Lukin, M. D.},
  journal   = {Nature},
  title     = {Entanglement of nanophotonic quantum memory nodes in a telecom network},
  year      = {2024},
  issn      = {1476-4687},
  month     = may,
  number    = {8012},
  pages     = {573--578},
  volume    = {629},
  doi       = {10.1038/s41586-024-07252-z},
  publisher = {Springer Science and Business Media LLC},
}

@Article{Kucera2024,
  author    = {Kucera, Stephan and Haen, Christian and Arenskötter, Elena and Bauer, Tobias and Meiers, Jonas and Schäfer, Marlon and Boland, Ross and Yahyapour, Milad and Lessing, Maurice and Holzwarth, Ronald and Becher, Christoph and Eschner, Jürgen},
  journal   = {npj Quantum Information},
  title     = {Demonstration of quantum network protocols over a 14-km urban fiber link},
  year      = {2024},
  issn      = {2056-6387},
  month     = sep,
  number    = {1},
  pages     = {88},
  volume    = {10},
  doi       = {10.1038/s41534-024-00886-x},
  publisher = {Springer Science and Business Media LLC},
}

@Article{McArdle2020,
  author    = {McArdle, Sam and Endo, Suguru and Aspuru-Guzik, Alán and Benjamin, Simon C. and Yuan, Xiao},
  journal   = {Reviews of Modern Physics},
  title     = {Quantum computational chemistry},
  year      = {2020},
  issn      = {1539-0756},
  month     = mar,
  number    = {1},
  pages     = {015003},
  volume    = {92},
  doi       = {10.1103/RevModPhys.92.015003},
  publisher = {American Physical Society (APS)},
}

@Article{Bharti2022,
  author    = {Bharti, Kishor and Cervera-Lierta, Alba and Kyaw, Thi Ha and Haug, Tobias and Alperin-Lea, Sumner and Anand, Abhinav and Degroote, Matthias and Heimonen, Hermanni and Kottmann, Jakob S. and Menke, Tim and Mok, Wai-Keong and Sim, Sukin and Kwek, Leong-Chuan and Aspuru-Guzik, Alán},
  journal   = {Reviews of Modern Physics},
  title     = {Noisy intermediate-scale quantum algorithms},
  year      = {2022},
  issn      = {1539-0756},
  month     = feb,
  number    = {1},
  pages     = {015004},
  volume    = {94},
  doi       = {10.1103/RevModPhys.94.015004},
  publisher = {American Physical Society (APS)},
}

@Article{Bartolucci2023,
  author    = {Bartolucci, Sara and Birchall, Patrick and Bombín, Hector and Cable, Hugo and Dawson, Chris and Gimeno-Segovia, Mercedes and Johnston, Eric and Kieling, Konrad and Nickerson, Naomi and Pant, Mihir and Pastawski, Fernando and Rudolph, Terry and Sparrow, Chris},
  journal   = {Nature Communications},
  title     = {Fusion-based quantum computation},
  year      = {2023},
  issn      = {2041-1723},
  month     = feb,
  number    = {1},
  pages     = {912},
  volume    = {14},
  doi       = {10.1038/s41467-023-36493-1},
  publisher = {Springer Science and Business Media LLC},
}

@Article{Tame2017,
  author    = {Dieleman, F. and Tame, M. S. and Sonnefraud, Y. and Kim, M. S. and Maier, S. A.},
  journal   = {Nano Letters},
  title     = {Experimental Verification of Entanglement Generated in a Plasmonic System},
  year      = {2017},
  issn      = {1530-6992},
  month     = nov,
  number    = {12},
  pages     = {7455--7461},
  volume    = {17},
  doi       = {10.1021/acs.nanolett.7b03372},
  publisher = {American Chemical Society (ACS)},
}

@Article{Vitali2007,
  author    = {Vitali, D. and Gigan, S. and Ferreira, A. and Böhm, H. R. and Tombesi, P. and Guerreiro, A. and Vedral, V. and Zeilinger, A. and Aspelmeyer, M.},
  journal   = {Physical Review Letters},
  title     = {Optomechanical Entanglement between a Movable Mirror and a Cavity Field},
  year      = {2007},
  issn      = {1079-7114},
  month     = jan,
  number    = {3},
  pages     = {030405},
  volume    = {98},
  doi       = {10.1103/PhysRevLett.98.030405},
  publisher = {American Physical Society (APS)},
}

@Article{Tchounda2023,
  author    = {Tchounda, S.R. Mbokop and Djorwé, P. and Engo, S.G. Nana and Djafari-Rouhani, B.},
  journal   = {Physical Review Applied},
  title     = {Sensor Sensitivity Based on Exceptional Points Engineered via Synthetic Magnetism},
  year      = {2023},
  issn      = {2331-7019},
  month     = jun,
  number    = {6},
  pages     = {064016},
  volume    = {19},
  doi       = {10.1103/PhysRevApplied.19.064016},
  publisher = {American Physical Society (APS)},
}

@Article{Massembele2024,
  author    = {Massembele, D. R. Kenigoule and Djorwé, P. and Sarma, Amarendra K. and Abdel-Aty, A.-H. and Engo, S. G. Nana},
  journal   = {Physical Review A},
  title     = {Quantum entanglement assisted via Duffing nonlinearity},
  year      = {2024},
  issn      = {2469-9934},
  month     = oct,
  number    = {4},
  pages     = {043502},
  volume    = {110},
  doi       = {10.1103/PhysRevA.110.043502},
  publisher = {American Physical Society (APS)},
}

@Article{Yang2020,
  author    = {Yang, Zhi-Bo and Liu, Jin-Song and Jin, Hua and Zhu, Qing-Hao and Zhu, Ai-Dong and Liu, Hong-Yu and Ming, Ying and Yang, Rong-Can},
  journal   = {Optics Express},
  title     = {Entanglement enhanced by Kerr nonlinearity in a cavity-optomagnonics system},
  year      = {2020},
  issn      = {1094-4087},
  month     = oct,
  number    = {21},
  pages     = {31862},
  volume    = {28},
  doi       = {10.1364/OE.404522},
  publisher = {Optica Publishing Group},
}

@Article{Djo2024,
  author    = {Djorwé, P. and Abdel-Aty, A.-H. and Nisar, K.S. and Engo, S.G.N.},
  journal   = {Optik},
  title     = {Optomechanical entanglement induced by backward stimulated Brillouin scattering},
  year      = {2024},
  issn      = {0030-4026},
  month     = dec,
  pages     = {172097},
  volume    = {319},
  doi       = {10.1016/j.ijleo.2024.172097},
  publisher = {Elsevier BV},
}

@Article{Rostand2025,
  author    = {Massembele, D.R.K. and Djorwé, P. and Emale, K.B. and Peng, Jia-Xin and Abdel-Aty, A.-H. and Nisar, K.S.},
  journal   = {Physica B: Condensed Matter},
  title     = {Low threshold quantum correlations via synthetic magnetism in Brillouin optomechanical system},
  year      = {2025},
  issn      = {0921-4526},
  month     = jan,
  pages     = {416689},
  volume    = {697},
  doi       = {10.1016/j.physb.2024.416689},
  publisher = {Elsevier BV},
}

@Article{Ghorbani2025,
  author    = {Ghorbani, N. and Motazedifard, Ali and Naderi, M. H.},
  journal   = {Physical Review A},
  title     = {Effects of quadratic optomechanical coupling on bipartite entanglements, mechanical ground-state cooling, and mechanical quadrature squeezing in an electro-optomechanical system},
  year      = {2025},
  issn      = {2469-9934},
  month     = jan,
  number    = {1},
  pages     = {013524},
  volume    = {111},
  doi       = {10.1103/PhysRevA.111.013524},
  publisher = {American Physical Society (APS)},
}

@Article{Li2023,
  author    = {Li, Zigeng and Li, Xiaomiao and Zhong, Xiaolan},
  journal   = {Optics Express},
  title     = {Optomechanical entanglement affected by exceptional point in a WGM resonator system},
  year      = {2023},
  issn      = {1094-4087},
  month     = may,
  number    = {12},
  pages     = {19382},
  volume    = {31},
  doi       = {10.1364/OE.488948},
  publisher = {Optica Publishing Group},
}

@Article{Lai2022,
  author    = {Lai, Deng-Gao and Chen, Ye-Hong and Qin, Wei and Miranowicz, Adam and Nori, Franco},
  journal   = {Physical Review Research},
  title     = {Tripartite optomechanical entanglement via optical-dark-mode control},
  year      = {2022},
  issn      = {2643-1564},
  month     = aug,
  number    = {3},
  pages     = {033112},
  volume    = {4},
  doi       = {10.1103/PhysRevResearch.4.033112},
  publisher = {American Physical Society (APS)},
}

@Article{Lai.2022,
  author    = {Lai, Deng-Gao and Liao, Jie-Qiao and Miranowicz, Adam and Nori, Franco},
  journal   = {Physical Review Letters},
  title     = {Noise-Tolerant Optomechanical Entanglement via Synthetic Magnetism},
  year      = {2022},
  issn      = {1079-7114},
  month     = aug,
  number    = {6},
  pages     = {063602},
  volume    = {129},
  doi       = {10.1103/PhysRevLett.129.063602},
  publisher = {American Physical Society (APS)},
}

@Article{DeJesus1987,
  author    = {DeJesus, Edmund X. and Kaufman, Charles},
  journal   = {Physical Review A},
  title     = {Routh-Hurwitz criterion in the examination of eigenvalues of a system of nonlinear ordinary differential equations},
  year      = {1987},
  issn      = {0556-2791},
  month     = jun,
  number    = {12},
  pages     = {5288--5290},
  volume    = {35},
  doi       = {10.1103/PhysRevA.35.5288},
  publisher = {American Physical Society (APS)},
}

@Article{Slim2025,
  author    = {Slim, Jesse J. and del Pino, Javier and Verhagen, Ewold},
  journal   = {Nature Communications},
  title     = {Programmable synthetic magnetism and chiral edge states in nano-optomechanical quantum Hall networks},
  year      = {2025},
  issn      = {2041-1723},
  month     = aug,
  number    = {1},
  pages     = {7471},
  volume    = {16},
  doi       = {10.1038/s41467-025-62541-z},
  publisher = {Springer Science and Business Media LLC},
}

@Article{Mathew2020,
  author    = {Mathew, John P. and Pino, Javier del and Verhagen, Ewold},
  journal   = {Nature Nanotechnology},
  title     = {Synthetic gauge fields for phonon transport in a nano-optomechanical system},
  year      = {2020},
  issn      = {1748-3395},
  month     = feb,
  number    = {3},
  pages     = {198--202},
  volume    = {15},
  doi       = {10.1038/s41565-019-0630-8},
  publisher = {Springer Science and Business Media LLC},
}

@Article{Brendel2017,
  author    = {Brendel, Christian and Peano, Vittorio and Painter, Oskar J. and Marquardt, Florian},
  journal   = {Proceedings of the National Academy of Sciences},
  title     = {Pseudomagnetic fields for sound at the nanoscale},
  year      = {2017},
  issn      = {1091-6490},
  month     = apr,
  number    = {17},
  pages     = {E3390},
  volume    = {114},
  doi       = {10.1073/pnas.1615503114},
  publisher = {Proceedings of the National Academy of Sciences},
}

@Article{Fang2017,
  author    = {Fang, Kejie and Luo, Jie and Metelmann, Anja and Matheny, Matthew H. and Marquardt, Florian and Clerk, Aashish A. and Painter, Oskar},
  journal   = {Nature Physics},
  title     = {Generalized non-reciprocity in an optomechanical circuit via synthetic magnetism and reservoir engineering},
  year      = {2017},
  issn      = {1745-2481},
  month     = jan,
  number    = {5},
  pages     = {465--471},
  volume    = {13},
  doi       = {10.1038/nphys4009},
  publisher = {Springer Science and Business Media LLC},
}

@Article{Adesso2007,
  author    = {Adesso, Gerardo and Illuminati, Fabrizio},
  journal   = {Journal of Physics A: Mathematical and Theoretical},
  title     = {Entanglement in continuous-variable systems: recent advances and current perspectives},
  year      = {2007},
  issn      = {1751-8121},
  month     = jun,
  number    = {28},
  pages     = {7821--7880},
  volume    = {40},
  doi       = {10.1088/1751-8113/40/28/S01},
  publisher = {IOP Publishing},
}

@Article{Adesso2006,
  author    = {Adesso, Gerardo and Illuminati, Fabrizio},
  journal   = {New Journal of Physics},
  title     = {Continuous variable tangle, monogamy inequality, and entanglement sharing in Gaussian states of continuous variable systems},
  year      = {2006},
  issn      = {1367-2630},
  month     = jan,
  pages     = {15--15},
  volume    = {8},
  doi       = {10.1088/1367-2630/8/1/015},
  publisher = {IOP Publishing},
}

@Article{Coffman2000,
  author    = {Coffman, Valerie and Kundu, Joydip and Wootters, William K.},
  journal   = {Physical Review A},
  title     = {Distributed entanglement},
  year      = {2000},
  issn      = {1094-1622},
  month     = apr,
  number    = {5},
  pages     = {052306},
  volume    = {61},
  doi       = {10.1103/PhysRevA.61.052306},
  publisher = {American Physical Society (APS)},
}

@Article{Plenio2005,
  author    = {Plenio, M. B.},
  journal   = {Physical Review Letters},
  title     = {Logarithmic Negativity: A Full Entanglement Monotone That is not Convex},
  year      = {2005},
  issn      = {1079-7114},
  month     = aug,
  number    = {9},
  pages     = {090503},
  volume    = {95},
  doi       = {10.1103/PhysRevLett.95.090503},
  publisher = {American Physical Society (APS)},
}

@Article{Vidal2002,
  author    = {Vidal, G. and Werner, R. F.},
  journal   = {Physical Review A},
  title     = {Computable measure of entanglement},
  year      = {2002},
  issn      = {1094-1622},
  month     = feb,
  number    = {3},
  pages     = {032314},
  volume    = {65},
  doi       = {10.1103/PhysRevA.65.032314},
  publisher = {American Physical Society (APS)},
}

@Article{Chen2025,
  author    = {Chen, Pengju and Luo, Da-Wei and Yu, Ting},
  journal   = {Physical Review Research},
  title     = {Optimal entanglement generation in optomechanical systems via Krotov control of covariance matrix dynamics},
  year      = {2025},
  issn      = {2643-1564},
  month     = feb,
  number    = {1},
  volume    = {7},
  doi       = {10.1103/PhysRevResearch.7.013161},
  publisher = {American Physical Society (APS)},
}

@Article{Chen.2025,
  author    = {Chen, Changling and Tang, Kai and Zhou, Yuxuan and Yi, KangYuan and Zhang, Xuan and Zhang, Xu and Guo, Haosheng and Liu, Song and Chen, Yuanzhen and Yan, Tongxing and Yu, Dapeng},
  journal   = {Physical Review Research},
  title     = {Hardware-efficient stabilization of entanglement via engineered dissipation in superconducting circuits},
  year      = {2025},
  issn      = {2643-1564},
  month     = apr,
  number    = {2},
  pages     = {L022018},
  volume    = {7},
  doi       = {10.1103/PhysRevResearch.7.L022018},
  publisher = {American Physical Society (APS)},
}
\end{document}